\documentclass[acmsmall]{acmart}
\usepackage{soul}
\definecolor{pumpkin}{RGB}{211, 84, 0}

\usepackage{graphicx}
\usepackage{arydshln}
\usepackage{xcolor}
\usepackage{tcolorbox}
\usepackage{hyperref}
\usepackage{enumitem}

\usepackage{refcount}

\definecolor{AutomatedAI}{HTML}{fffbc0}
\definecolor{IntermediateAI}{HTML}{dcdff1}
\definecolor{MinimalAI}{HTML}{fdd4e2}
\newcommand{\AutomatedAI}{\texttt{\textcolor{black}{\fcolorbox{AutomatedAI}{AutomatedAI}{Automated AI}} }}
\newcommand{\IntermediateAI}{\texttt{\textcolor{black}{\fcolorbox{IntermediateAI}{IntermediateAI}{Intermediate AI}} }}
\newcommand{\MinimalAI}{\texttt{\textcolor{black}{\fcolorbox{MinimalAI}{MinimalAI}{Minimal AI}} }}

\AtBeginDocument{%
  }

\setcopyright{cc}
\setcctype{by}

\acmJournal{PACMHCI}
\acmYear{2025} \acmVolume{9} \acmNumber{7} \acmArticle{CSCW451}
\acmMonth{11} \acmDOI{10.1145/3757632}

\author{Xinyue Chen}
\email{xinyuech@umich.edu}
\orcid{0000-0002-0774-223X}
\affiliation{%
  \institution{University of Michigan}
  \city{Ann Arbor}
  \state{Michigan}
  \country{USA}
}

\author{Kunlin Ruan}
\email{kunlinvs@umich.edu}
\orcid{0009-0001-9457-5467}
\affiliation{%
  \institution{University of Michigan}
  \city{Ann Arbor}
  \state{Michigan}
  \country{USA}
}

\author{Kexin Phyllis Ju}
\email{kexinju@umich.edu}
\orcid{0009-0002-8272-6552}
\affiliation{%
  \institution{University of Michigan}
  \city{Ann Arbor}
  \state{Michigan}
  \country{USA}
}

\author{Nathan Yap}
\email{nyap@umich.edu}
\orcid{0009-0003-7760-1496}
\affiliation{%
  \institution{University of Michigan}
  \city{Ann Arbor}
  \state{Michigan}
  \country{USA}
}

\author{Xu Wang}
\email{xwanghci@umich.edu}
\orcid{0000-0001-5551-0815}
\affiliation{%
  \institution{University of Michigan}
  \city{Ann Arbor}
  \state{Michigan}
  \country{USA}
}
\received{October 2024}
\received[revised]{April 2025}
\received[accepted]{August 2025}
\begin{document}

\title[AI Assistance Dilemma in AI-Supported Note-Taking]{More AI Assistance Reduces Cognitive Engagement: Examining the AI Assistance Dilemma in AI-Supported Note-Taking}



\begin{abstract}
As AI tools become increasingly embedded in cognitively demanding tasks such as note-taking, questions remain about whether they enhance or undermine cognitive engagement. This paper examines the "AI Assistance Dilemma" in note-taking, investigating how varying levels of AI support affect user engagement and comprehension. In a within-subject experiment, we asked participants (N=30) to take notes during lecture videos under three conditions: \AutomatedAI (high assistance with structured notes), \IntermediateAI (moderate assistance with real-time summary, and \MinimalAI (low assistance with transcript). Results reveal that Intermediate AI yields the highest post-test scores and Automated AI the lowest. Participants, however, preferred the automated setup due to its perceived ease of use and lower cognitive effort, suggesting a discrepancy between preferred convenience and cognitive benefits. Our study provides insights into designing AI assistance that preserves cognitive engagement, offering implications for designing moderate AI support in cognitive tasks.
\end{abstract}



\begin{teaserfigure}
  \includegraphics[width=\textwidth]{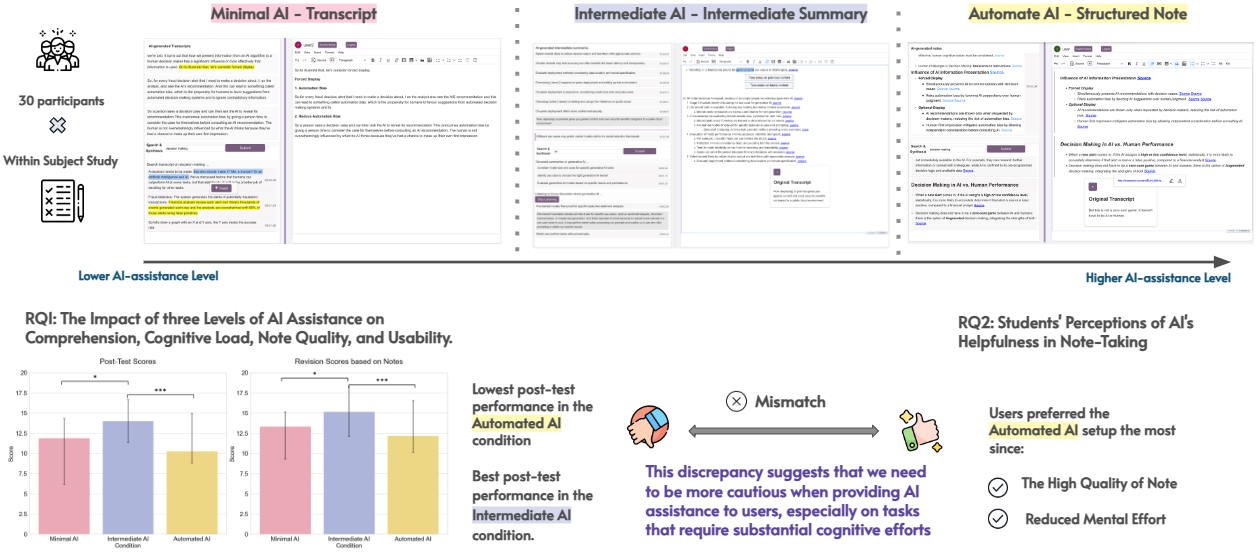}
  \caption{In a within-subject experiment with 30 participants, we tested three variants of an AI-assisted note-taking system to help people understand the content of a live lecture: \AutomatedAI (high assistance), \IntermediateAI (moderate assistance with summaries), and \MinimalAI (transcript-only). Results show \IntermediateAI yielded the highest learning outcomes, while \AutomatedAI scored lowest despite being the most preferred for ease of use. These findings highlight the gap between preferred convenience and cognitive benefit, offering guidance for designing AI that enhances cognitive engagement in learning tasks.}
  \label{fig:teaser}
\end{teaserfigure}

\maketitle


\section{Introduction}

AI is increasingly used to facilitate cognitively challenging tasks, from providing next-sentence suggestions in writing \cite{dhillon2024shaping} to generating meeting summaries \cite{asthana2023summaries, chen2023meetscript, wangMeetingBridgesDesigning2024}. The rise of AI productivity tools \cite{shaer_ai-augmented_2024, schroder_autoscrum_2023, gao2023coaicoder} raises the question: is AI assistance always desired and beneficial? When AI is leveraged to enhance productivity, could it compromise people's cognitive engagement in the task, leading to suboptimal outcomes? In this paper, we examine the AI assistance dilemma in the context of a crucial and prevalent task: note-taking. We examine to what extent AI-powered automatic note-taking services might reduce valuable cognitive engagement of the note-taker and how we could design human-AI collaborative mechanisms to provide the necessary AI assistance to the note-taker while preserving the cognitive effort required for meaningful content engagement.


Note-taking is a common activity during real-time information consumption, such as in meetings \cite{umapathy_requirements_nodate, teleniusSensemakingMeetingsCollaborative2016}, lectures \cite{rongUnderstandingPersonalData2023, goyalEffectsVisualizationNotetaking2013}, and presentations, where participants must quickly process information without the ability to revisit what was said \cite{chen2023meetscript, kuzminykhClassificationFunctionalAttention2020}. Note-taking has proven to be a helpful tool to assist real-time information consumption by externalizing information and reducing the load on participants' working memory 
\cite{kuzminykhClassificationFunctionalAttention2020}. Note-taking serves two key functions: it helps people internalize knowledge through active engagement (encoding function) and provides a lasting reference for future recall(storage function) \cite{kiewra_note-taking_1991, kiewraReviewNotetakingEncodingstorage1989}.





Note-taking offers a unique and valuable context to study the AI assistance dilemma due to its dual demands on cognitive processing and external documentation \cite{witherby_current_2019, flaniganImpactDigitalDistraction2020}. Effective note-taking requires users to engage deeply with information, selecting and organizing key points in real-time, which promotes active learning and understanding \cite{makanyOptimisingUseNote2009, fang_understanding_2022}. However, when AI tools are introduced to automate or aid in this process, there is potential for reduced cognitive engagement, as users may rely on the AI rather than actively process information themselves \cite{huq2024noteeline}. 

It thus raises important questions about AI’s role: Does AI assistance merely alleviate cognitive load, or does it undermine the mental engagement crucial for comprehension and retention?
This concern reflects a longstanding issue known as the "Assistance Dilemma" in learning sciences, originally identified in the context of cognitive tutors deployed in classrooms \cite{koedingerExploringAssistanceDilemma2007}. This dilemma underscores that while AI-generated feedback can enhance learning, excessive assistance risks undermining learners' capacity for independent learning and critical thinking, whereas insufficient assistance may make tasks overly challenging \cite{koedingerExploringAssistanceDilemma2007, roll2007designing}. 
With the growing prevalence of AI assistance, we aim to investigate to what extent the AI assistance dilemma exists and how we might design human-AI collaborative mechanisms to address this challenge in a note-taking context.  

To this end, we designed three levels of AI assistance to help people take notes during live lectures,
namely 1) \AutomatedAI, where participants received auto-generated AI notes around every 90 seconds, similar to common note-taking services; 2) \IntermediateAI, where participants received real-time AI summary blocks; and 3) \MinimalAI, where participants received real-time transcript blocks. In all three conditions, participants take notes in a text editor. Participants can drag and drop the AI-generated information (structured notes, summary blocks, and transcript blocks) into the text editor to compose their own notes. 
The \AutomatedAI setup resembles popular note-taking services that provide a summary after a time interval \footnote{http://otter.ai/}. With this setup, we aim to answer whether automated note-taking support would compromise people's cognitive engagement and negatively impact their understanding of the lecture content. 
By introducing the \IntermediateAI and the \MinimalAI setups, we aim to address whether giving users more control and agency during the note-taking process would enhance their understanding of the content compared to the \AutomatedAI setup. 

\begin{figure}
    \centering
    \includegraphics[width=0.9\linewidth]{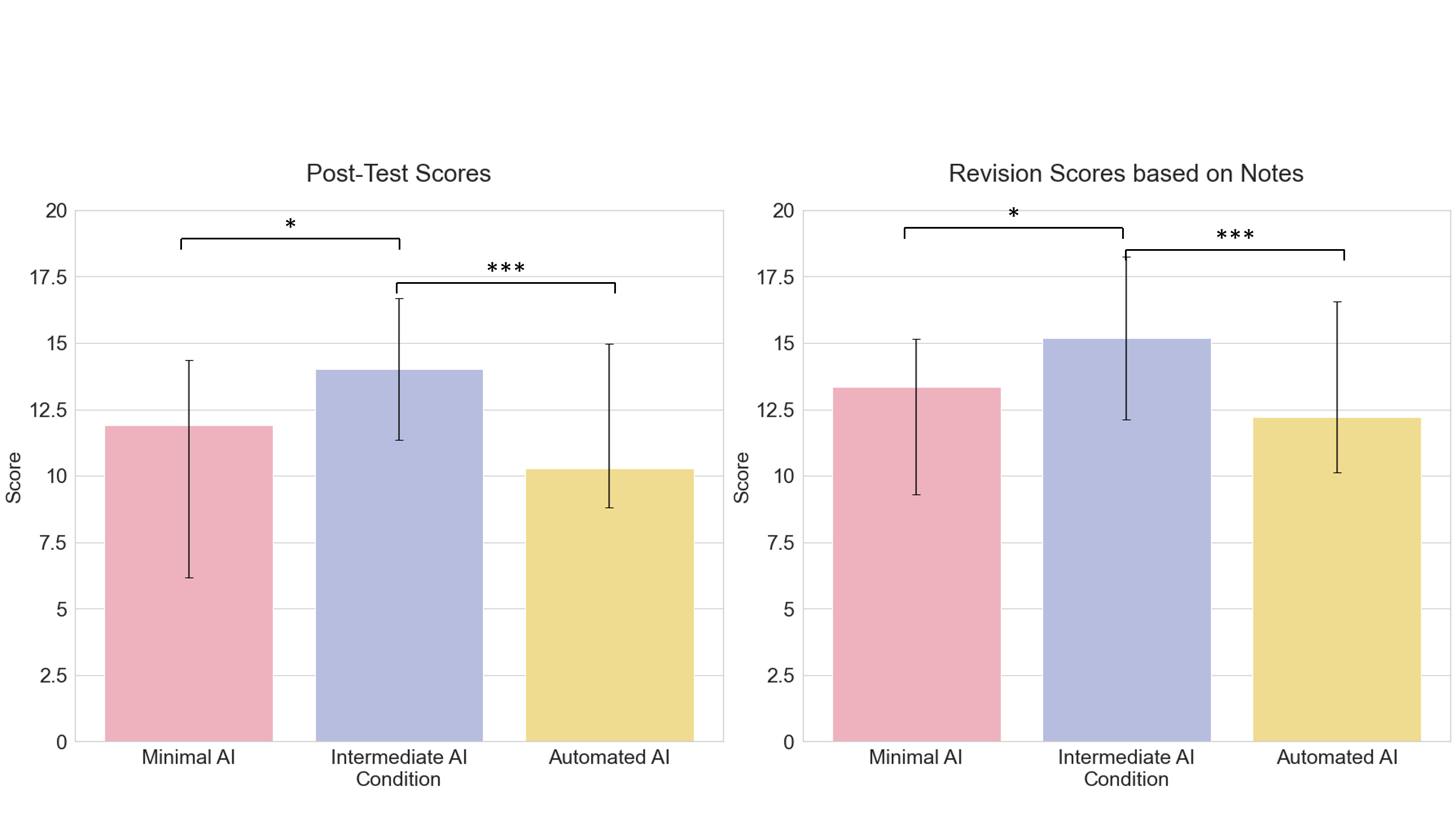}
    \caption{\textbf{Post-hoc Tukey HSD analysis comparing post-test and revision scores across conditions.} The left bar graph shows the \textsc{Post-test Scores}, where the \IntermediateAI condition significantly outperformed both the \AutomatedAI and \MinimalAI. The right bar graph displays the \textsc{Revision Scores}, with similar trends observed.
    }
    \label{post-test}
\end{figure}

We performed a within-subject experiment with 30 participants, where they watched lecture videos and took notes in the three setups. 
Our findings suggest that 1) the AI assistance dilemma does exist. People had the lowest post-test performance in the \AutomatedAI condition and the best post-test performance in the \IntermediateAI condition. The difference is statistically significant (Figure \ref{post-test}). 
This finding suggests that although AI may enhance productivity, it risks compromising people's cognitive engagement and may lead to lower comprehension and retention. 2) Providing intermediate summary blocks is a useful way to address this challenge, as users in the \IntermediateAI condition had the best post-test performance. Users had more control and more valuable cognitive engagement during the note-taking process when they were provided with the building blocks and had the agency to compose notes by themselves. 
 3) However, users preferred the \AutomatedAI setup the most. They found that the auto-generated notes were of high quality and that they reduced their effort significantly.  These findings highlight a discrepancy between preferred convenience and cognitive benefit. 
 This discrepancy suggests that we need to be more cautious when providing AI assistance to users, especially on tasks that require substantial cognitive efforts. As users naturally prefer to minimize effort, the risk of over-reliance on AI becomes even more pronounced.

\section{Related Work}

\subsection{Cognitive Costs and Benefits of Note-taking}

Lecture note-taking has long been shown to be effective for learning. In-class note-taking helps students actively process information \cite{makanyOptimisingUseNote2009}, improving memory retention \cite{fanguyHowCollaborationInfluences2023}, while the notes can also serve as an external representation for post-class review, which reinforces understanding and supports long-term knowledge retention \cite{piolatCognitiveEffortNote2005b, makanyOptimisingUseNote2009}. The underlying mechanism for these effects, known as the encoding-storage paradigm \cite{kiewraReviewNotetakingEncodingstorage1989}, suggests two key functions: \textsc{encoding}, where recording and taking notes itself promotes learning through increased attention and deeper processing, and \textsc{storage}, where notes act as external memory for later review, reinforcing understanding by consolidating information and helping to slow forgetting.

Many studies have
shown detrimental effects of taking notes while learning, as note-taking costs attentional resources that are needed for processing rapid and dense lecture presentations \cite{buiNotetakingComputersExploring2013}. Research connecting note-taking with cognitive load theory suggests that when students attempt to write down information verbatim, it can lead to an increase in extraneous cognitive load, which negatively impacts learning effectiveness \cite{wongTakeNotesNot2023}. 
To optimize note-taking, researchers have explored structured techniques, finding that formats like pre-organized sheets \cite{fang_understanding_2022} and outlines \cite{kiewra_note-taking_1991} reduce extraneous load and promote germane processes such as comprehension and synthesis \cite{costley2021collaborative, jansen2017integrative}. These structured approaches can also enhance metacognition, encouraging students to focus on understanding knowledge structures and how to prioritize, organize, and learn effectively \cite{fang_understanding_2022}. 

In summary, past research underscores the importance of reducing unnecessary cognitive load while supporting note-taking’s encoding and storage functions. Building on these theories, our study explores how varying levels of AI assistance impact encoding, storage, and cognitive load optimization in note-taking.

\subsection{Tools to Enhance Note-taking}
HCI researchers have developed numerous tools to support note-taking. One line of the research focuses on note-taking modalities and input methods, offering flexibility for capturing notes via voice \cite{khanTypeSpeakEffect2022}, digital ink \cite{hinckley2007inkseine}, sketching \cite{zheng2021sketchnote},  and mobile inputs \cite{ren2014inkanchor}. Studies comparing note-taking modalities reveal that while typing enables faster capture, handwriting better supports encoding due to its hands-on nature, though results vary across contexts \cite{bauer2007selection, flaniganImpactDigitalDistraction2020}. While these tools emphasize input affordances, they didn't address real-time, content-specific support within classroom settings. 
Researchers also seek ways to scaffold real-time classroom note-taking, including guiding students to take notes in structured formats or with predefined structures, such as concept maps \cite{sunHowStudentsGenerate2022, chenAssociationReflectionStimulation2021} or designing collaborative tools such as NotePals \cite{davis1999notepals}, LiveNotes \cite{kam2005livenotes}, or NoteCoStruct \cite{fang2021notecostruct} that reduce individual cognitive load by distributing note-taking responsibilities among group members. While collaborative notes are proven to be of higher quality \cite{liu2019notestruct}, they may compromise individual agency since students lean towards relying on contributions made by their peers instead of personal involvement \cite{costley2021collaborative, fanguyHowCollaborationInfluences2023, costley2022interaction}. This dependence can weaken the personal encoding benefits of note-taking, thus reducing comprehension and retention \cite{costley2021collaborative}.

Recently, AI has increasingly supported automated note generation, especially in complex contexts like clinical documentation \cite{leong2024gen, wang2022phenopad, tsai2024gazenoter} and online meetings \cite{chen2023meetscript}. 
While AI automation is advantageous in environments where efficiency is paramount, classroom note-taking differs in that it serves not only as information storage but as a critical encoding process \cite{kiewraReviewNotetakingEncodingstorage1989}. Fully automated notes in classrooms may inhibit the encoding process, yet research on the ideal level of AI assistance in note-taking remains scarce. While a few recent studies started exploring using AI to generate an intermediate level of notes that users can further work on, providing insights into balancing user engagement and AI assistance \cite{chen2025meetmap}, the impact of AI on real-time encoding and post-lecture storage or decoding processes in educational settings has yet to be systematically quantified and understood.

\subsection{AI Assistance Dilemma in Cognitive Tasks} 

The assistance dilemma describes the challenge of balancing assistance with autonomy in learning, highlighting the need to provide optimal levels of support without fostering dependency \cite{koedingerExploringAssistanceDilemma2007}. In educational research, solutions to the assistance dilemma emphasize carefully calibrated scaffolding strategies to guide students' meta-cognitive skills \cite{roll2011improving} and adjust the level of help based on learners' skill and engagement levels \cite{kalyuga2009adapting}. For example, in early problem-solving stages, worked examples are frequently paired with minimal guidance, transitioning to increased feedback or step-by-step hints as learners advance, encouraging their independent reasoning \cite{mclaren2008and}. 

In cognitive tutoring, studies indicate that varying levels of scaffolding are effective, yet these studies primarily pertain to predefined problem domains and structured worked examples, where they can effectively trace students' knowledge progression \cite{mclaren2008and, wood2001scaffolding}.  This work extends the understanding of the AI assistance dilemma beyond structured learning environments by examining how different levels of AI support affect comprehension and engagement in the context of real-time note-taking, which involves open-ended, temporally constrained sense-making.

\subsection{Design to Balancing AI assistance in Sense-making Activities} \label{insigts}
Recent research on using large language models (LLMs) for sense-making tasks in diverse scenarios provides insights into how to balance AI assistance and human engagement in cognitive and sense-making tasks \cite{dhillon2024shaping,mcnuttDesignAIpoweredCode2023, chen2025meetmap, lin2024rambler, zhang2023visar}. 

Prior work has shown that AI can support user engagement more effectively when it provides partial assistance rather than completing the task entirely \cite{dhillon2024shaping, mcnuttDesignAIpoweredCode2023}. For example, in writing tasks, offering full paragraph suggestions may reduce user agency, while sentence-level assistance effectively reduces the difficulty of phrasing without limiting creative freedom \cite{dhillon2024shaping}. 

People also explored interactive mechanisms to help people engage more in sense-making activities when receiving AI assistance.  For instance, in the MeetMap system, researchers introduced a temporary AI-generated content holding area, allowing users to decide which suggestions to use or ignore \cite{chen2025meetmap}. This encourages users to reflect on the content before accepting it, which can help maintain cognitive engagement \cite{chen2025meetmap, zhang2023visar}. 

Additionally, the level of AI abstraction and its potential uncertainty are important design considerations \cite{gao2023coaicoder, gu2024ai}. For instance, a resilient interface can continuously highlight critical information, increasing AI’s abstraction of original content as needed \cite{gu2024ai}.

The design of the three levels of AI assistance in this study draws insights from these designs, balancing the granularity of AI-provided content and the agency provided to users, using a holding area where users can further interact with the AI-generated content, and adjusting abstraction levels to support effective note-taking.

\section{Study}
As demonstrated in previous research, note-taking is vital for real-time understanding and sense-making during lectures, yet students often lack effective techniques, reducing its cognitive benefits. AI tools have been introduced to aid note-taking, but fully automated methods may undermine the very cognitive processes that make it valuable. This tension - known as the assistance dilemma, has been well studied in structured learning environments such as cognitive tutoring. However, its implications for real-time, open-ended sense-making activities like lecture note-taking remain underexplored.

To bridge this gap, we aim to investigate to what extent the AI assistance dilemma exists and how we might design human-AI collaborative mechanisms to address this challenge in a note-taking context. We address these questions through the design and evaluation of a system that supports three distinct levels of AI assistance, namely \AutomatedAI - which provides slightly delayed AI-generated structured notes every 1-2 minutes, \IntermediateAI - which presents real-time summary blocks, and \MinimalAI - which provides real-time transcript blocks.

We performed a within-subject experiment with 30 participants, where they watched lecture videos and took notes in the three setups.  We introduce a post-test of the lecture to measure people's understanding, which is a signal of their cognitive engagement level during the lecture. We also gave the participants an opportunity to revise their post-test answers based on the notes, which is a signal of how much information they could retrieve from the notes they took. 



\subsection{Experiment design}

\subsubsection{Apparatus}
 We designed the note-taking tasks with three distinct levels of AI assistance, drawing from prior research on human-AI collaboration tools for sense-making activities, as discussed in section \ref{insigts}. In those studies, the assistance level was differentiated based on the degree of abstraction and scaffolding provided by AI, as well as the level of user agency retained, emphasizing how manual effort is distributed between humans and AI \cite{dhillon_shaping_2024, gu2024ai}. In the context of lecture note-taking in this research, the assistance levels reflect how much of the encoding process AI automates and the degree of user autonomy in managing and engaging with content. Besides, the level of abstraction in such real-time tasks may also be related to the synchronicity users receive the AI assistance \cite{lin2024rambler}. 
In this study, the assistance level apparatus is designed as follows: 
\begin{itemize}

\item \AutomatedAI: AI generates a structured note block around every two minutes. Users can incorporate these structured blocks into their own notes - the notes are shown in a granularity that is similar to the business usual AI note-taking tools \footnote{https://www.zoom.com/en/ai-assistant/}\footnote{https://support.microsoft.com/en-us/office/use-copilot-in-microsoft-teams-meetings}.
\item  \IntermediateAI: AI produces turn-level summary blocks for each speaking turn. Users can select and incorporate these intermediate summary blocks into their notes.
 \item \MinimalAI: AI provides a real-time transcript divided into transcript blocks after each completed speaking turn. Users can select and incorporate these transcript blocks into their notes. 
    
\end{itemize}

The detailed system features and how the AI assistance differs in the three setups will be discussed in section \ref{3.2}


\subsubsection{Participants and Procedure}
We recruited 30 participants through mailing lists at the University of Michigan. We selected participants who indicated that they faced challenges in note-taking. The demographic information of participants is shown in Table \ref{tab:demographic}.

\begin{table}[h!]
\centering

\begin{tabular}{lcclc}
\hline
\textbf{Gender} & & & \textbf{Age} & \\
\cline{1-1} \cline{4-4}
Male & 16 & & 0 - 20 years old & 9 \\
Female & 9 & & 21 - 30 years old & 21 \\
Genderqueer & 1 & & & \\
Prefer not to say & 4 & & & \\
\hline
\textbf{Ethnicity} & & & \textbf{Year in School} & \\
\cline{1-1} \cline{4-4}
Caucasian & 7 & & Freshman & 0 \\
Asian & 19 & & Sophomore & 9 \\
Other/Unknown & 4 & & Junior & 4 \\
& & & Senior & 12 \\
& & & Master & 1 \\
& & & PhD & 4 \\
\hline
\end{tabular}
\caption{Demographic Information of Participants}
\label{tab:demographic}
\end{table}

\begin{figure}
    \centering
    \includegraphics[width=1\linewidth]{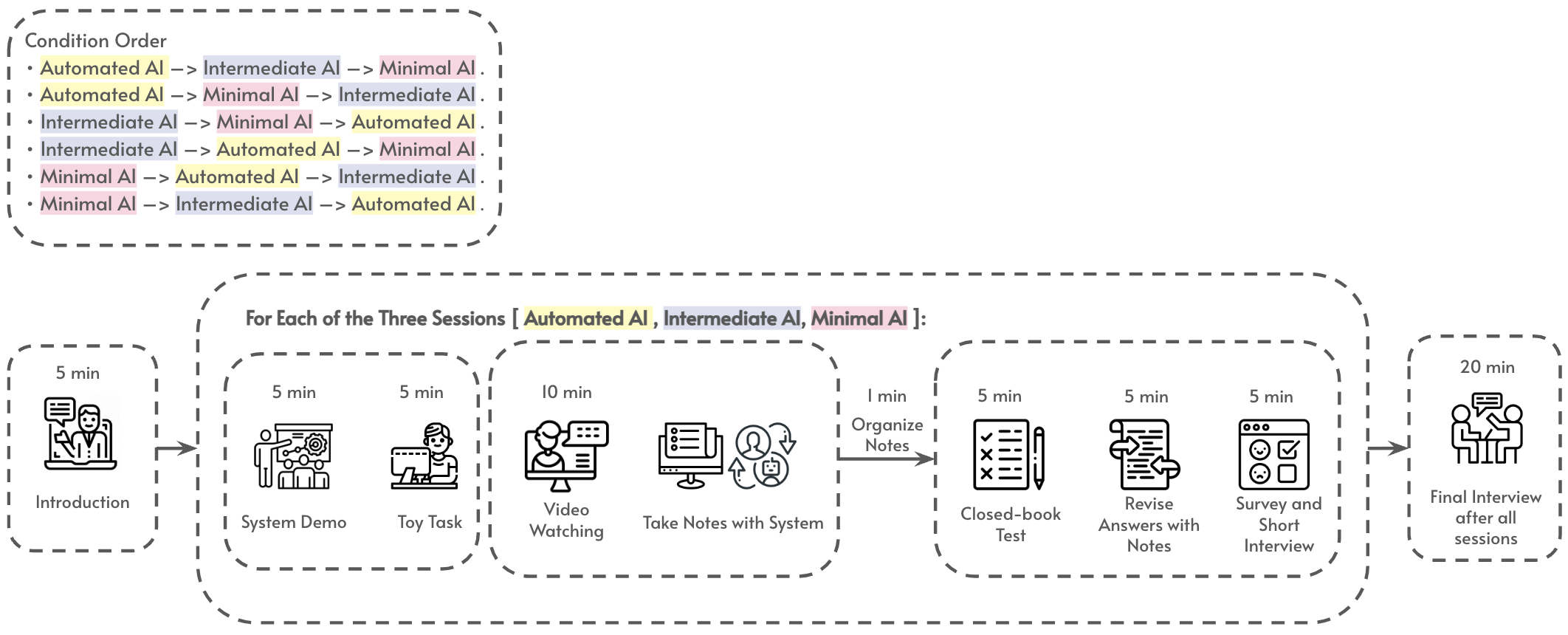}
    \caption{\textbf{Study Design.} The study started with a 5-minute introduction outlining its objectives. Participants then experienced all three AI assistance conditions— \AutomatedAI, \IntermediateAI, \MinimalAI, in a specified order. Each session began with a demo instruction and a toy task. Participants watched a 10-minute video while taking notes with the system, after which they had 1 minute to organize their notes. This was followed by a closed-book test lasting 5 minutes, after which participants revised their answers with their notes for an additional 5 minutes. Each session concluded with a survey and a brief interview lasting 5 minutes. A final interview was conducted after all sessions, lasting 20 minutes.}
    \label{fig:study_design}
\end{figure}

Each participant experienced all three AI assistance conditions in person, i.e., \AutomatedAI, \IntermediateAI, \MinimalAI. 
The order of the conditions is counterbalanced across participants to mitigate order effects.
Participants were randomly assigned to one of six possible sequences to ensure balanced exposure, as shown in Figure \ref{fig:study_design}.

The experiment began with an introduction to the study, providing participants with a thorough explanation of the tasks. A demo was played for each system variant, after which participants engaged in a toy task designed to familiarize them with the interface and functionalities of each system.

For each task, participants were seated in a study room with a large display simulating a classroom setting. The display played a 10-minute lecture video, mimicking a live classroom environment where the instructor delivered the lecture. During this time, participants used their own computers to take notes in real time through the customized AI-assisted system. To replicate the demands of in-situ note-taking, the video was designed to be non-replayable.

At the end of each video, participants were given one minute to finalize their notes before proceeding to a post-test, which included two multiple-choice questions (MCQs) and three open-ended questions (OEQs) \footnote{ \label{resource} Links to the study materials: https://drive.google.com/file/d/17mWKqU3ydLI-dtcm2R-iwwrN7wq9rxdw }. Initially, participants completed the test without referring to their notes to assess the influence of the encoding process on learning outcomes \cite{kiewraReviewNotetakingEncodingstorage1989}. They then submitted their responses and were asked to revisit and revise their answers with access to their notes, aligning with prior research on the storage paradigm, where notes serve as external storage for review and reinforcement \cite{kiewraReviewNotetakingEncodingstorage1989}.
Following the post-test, participants completed a post-task survey assessing their cognitive load and user experience with the system. Additionally, each task concluded with a 10-minute interview to capture immediate feedback and insights on the specific condition they had just completed.

Upon completion of all three tasks, participants participated in a final 20-minute comprehensive interview to discuss their comparative perceptions of the three AI assistance levels. Questions were tailored to explore participants' interactions with intermediate and structured AI-generated notes, their cognitive demands and agency during the process, and the trustworthiness of the AI.
In total, the study took approximately two hours per participant. 

\subsubsection{Material} 
We selected three lecture videos from YouTube, all presented by the same instructor and centered on an engaging, timely topic: large language models (LLMs). Each video spans approximately 8 minutes and 50 seconds to 9 minutes and 30 seconds. The topics covered are "Introduction to AI-Augmented Decision-Making," "Introduction to LLM Hallucination," and "Introduction to LLM model selection framework for your task," with the order fixed as listed.  Importantly, the videos required no prior knowledge. Careful consideration was given to the selection and balance of video length and topic, minimizing potential differences in difficulty or familiarity and reducing these variables' impact on learning outcomes. 

For each video, we designed a post-test comprising two multiple-choice questions (MCQs) and three open-ended questions (OEQs). An expert instructor in AI developed the tests, and their quality was validated in a pilot study involving four students who watched the videos, answered the questions, and provided feedback on any points of confusion. This feedback ensured that the questions were clear and answerable based solely on the lecture content without needing outside resources. Full details on video sources and test questions can be found in the supplementary materials \footnotemark[\getrefnumber{resource}].

We opted not to include a pre-test in this study because the selected lecture videos are closely matched in length, complexity, and topic, minimizing video difficulty as a factor in learning outcomes. 
We excluded it to avoid lengthening the study since students had already experienced three conditions. 

\subsection{NoteCopilot: The Custom-Built AI-assisted Note-taking Tools}
\label{3.2}
We developed NoteCopilot, an AI-assisted note-taking system offering adjustable levels of AI assistance - \MinimalAI, \IntermediateAI, \AutomatedAI, each tailored to examine different degrees of cognitive assistance in note-taking. 

\subsubsection{Common interaction across the three system variants}

\begin{figure}
    \centering
    \includegraphics[width=1\linewidth]{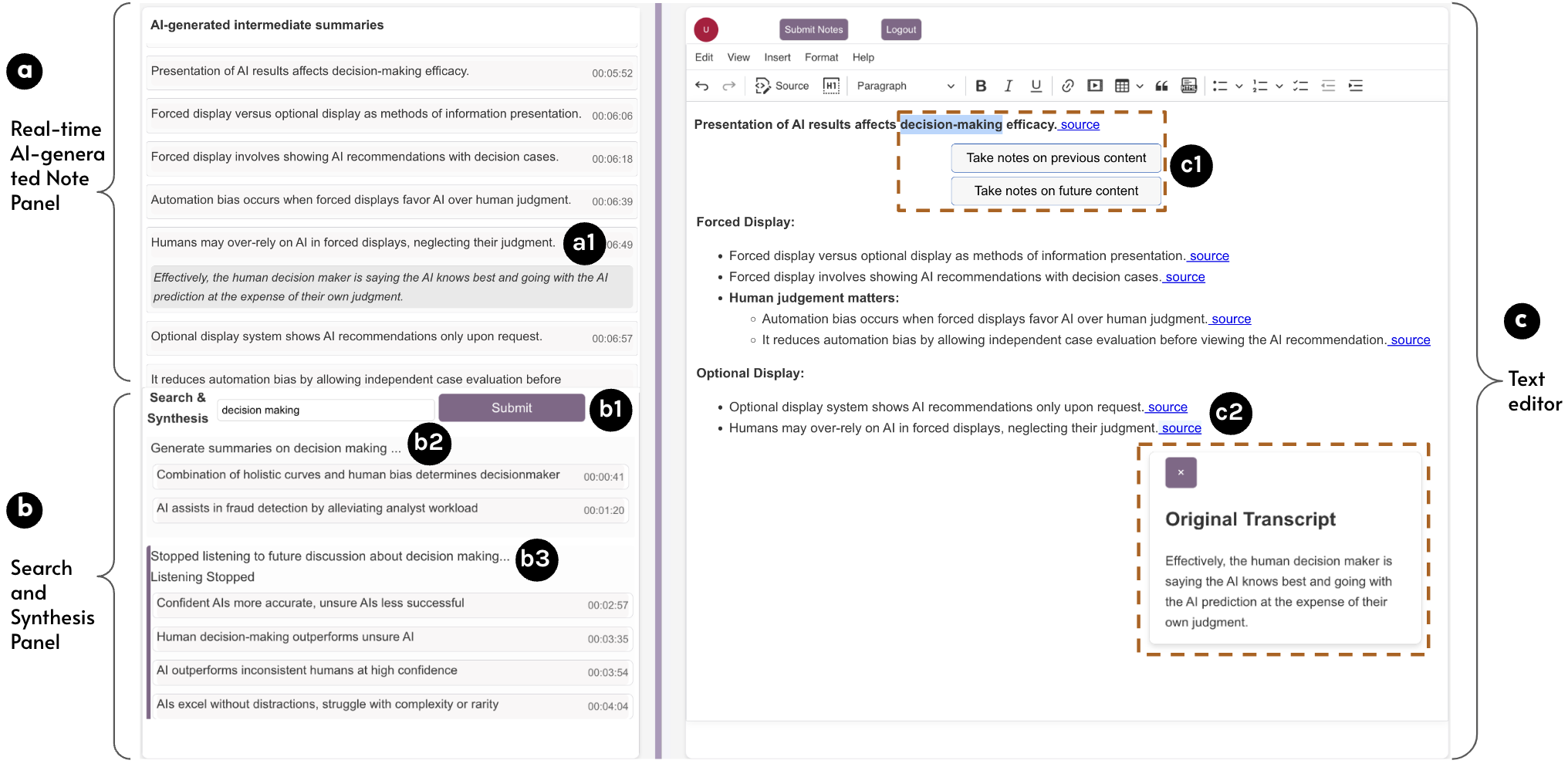}
    \caption{\textbf{System Overview.} On the upper left is the \textsc{Real-time AI-Generated Note} panel (a), where users can see AI-generated notes and click on them to see related transcripts (a1). On the lower left is the \textsc{Search and Synthesis Panel} (b), where users can interact with AI by requesting notes centered around a topic (b1) based on previous course content.(b2). On the right is the rich Text-Editor (c), where users can create their own notes or, with the help of AI-generated notes, drag and drop the AI content from (a) or (b) to the editor. They can use in-context selection (c1) to request topic-related notes based on previous course content as (b1) does and also ask AI to continuously generate topic-related notes by listening to future course content and receive results as (b3). In addition, users can check original related transcripts for each AI-generated note by clicking on the source button (c2). This is the interface for \IntermediateAI. The interfaces for \AutomatedAI and \MinimalAI follow a similar structure, except that the note blocks in all panels are larger structured summaries or transcripts, respectively.}
    \label{fig:system}
\end{figure}

The interface includes a \textsc{Real-Time AI-generated Note Panel} for proactive assistance, a \textsc{Search and Synthesize Panel} for expressing specific user intents and retrieving personalized results, and a \textsc{rich text editor}. 
An overview of the interface is shown in Figure \ref{fig:system}.

The \textsc{Real-Time AI-generated Note Panel} (Figure \ref{fig:system}(a)) displays AI-generated notes in real-time. Depending on the condition, the content in this panel varies from full transcripts (in \MinimalAI) to intermediate summaries (in \IntermediateAI) and structured notes (in \AutomatedAI). This feature enables users to view AI-generated content as the lecture progresses, serving as the system’s core AI assistance. Here, AI proactively provides content in a temporary middleware layer, allowing users to opt to integrate it into their notes or not. This design aligns with prior research that recommends a temporary holding space for AI content rather than direct insertion into the editor \cite{chen2025meetmap} as discussed in \S \ref{insigts}.

The \textsc{text editor} (Figure \ref{fig:system}(c)) offers full note-taking functionality, supporting typing, selection, editing, and multimedia formatting. Users can drag and drop AI-generated content from panels (a) and (b) into the editor (c). When AI-generated content is dragged into the editor, a source button (c2) allows users to view the original transcripts associated with AI summaries, ensuring transparency and traceability of the information.

The above two panels form the system’s core mechanism, where AI generates notes in real-time with varying granularity and timing, allowing students the flexibility to selectively incorporate AI-generated content into their notes.

In addition to proactive AI assistance, we offer users the flexibility to search for specific information as needed. The \textsc{Search and Synthesis Panel}  (Figure \ref{fig:system}(b)) enables users to interact with the AI by requesting targeted information on lecture topics. Users can submit a query (b1), and the AI provides responses that are aligned with the AI assistance levels based on their request (b2). Additionally, users can initiate searches directly within the editor (c). By selecting typed content, users can access a drop-down menu with two AI options: \textit{"Take notes on previous content"} and \textit{"Take notes on ongoing content"} (c1). The former generates AI notes based on previous lecture content, while the latter activates a listening mode, capturing any mentions of the selected topic in future content. Importantly, the granularity and abstraction of the AI responses in this panel correspond to the assigned condition: in the \MinimalAI condition, search results return transcript blocks; in the \IntermediateAI condition, they return summary blocks; and in the \AutomatedAI condition, they provide structured notes. 

The interaction design remains consistent across the three assistance levels, enabling users to interact with the AI panels similarly regardless of the specific condition. This consistency allows us to focus on examining how varying levels of AI processing and automation impact user engagement and learning.

\subsubsection{Three AI-assistance Levels}
While users interact with AI in a similar way, each system variant—\MinimalAI, \IntermediateAI, and \AutomatedAI—offers unique levels of AI assistance.  
The three assistance levels reflect different extents to which AI automates the encoding process—the cognitive effort of interpreting, structuring, and recording information, and to what extent AI abstracts the information and adds its own interpretation to the content.

\begin{figure}
    \centering
    \includegraphics[width=1\linewidth]{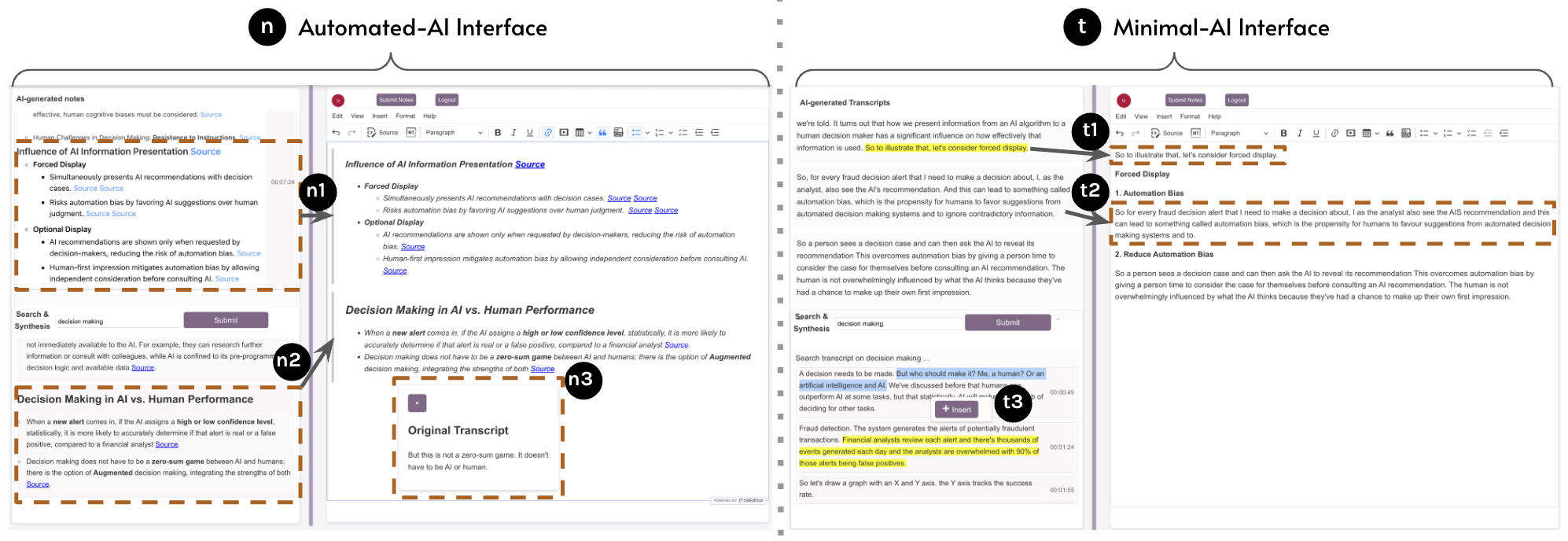}
    \caption{\textbf{\AutomatedAI and \MinimalAI Interface.} On the left is the \textsc{\AutomatedAI Interface} (n), where users can drag and drop AI-generated notes alongside related transcripts from the \textsc{Real-time AI-Generated Note panel} (n1) and \textsc{Search and Synthesis Panel} (n2) into the \textsc{Text-Editor Panel}. Users can also expand related transcripts by clicking the source button (n3). On the right is the \textsc{\MinimalAI Interface} (t), allowing users to directly select text from the \textsc{Real-time AI-Generated Note panel} (t1) and \textsc{Search and Synthesis Panel} (t3) for highlighting and insertion into the \textsc{Text-Editor Panel}. Additionally, users can drag and drop entire transcript blocks into the \textsc{Text-Editor Panel} (t2).}
    \label{fig:note_transcript}
\end{figure}

 In the\AutomatedAI condition, AI generates structured note blocks approximately every 1–2 minutes (Figure~\ref{fig:note_transcript}(n1)). This interval is chosen to provide users with timely content while allowing AI sufficient time to process and organize content into coherent structures. 
 This design was assumed to reduce the users’
note-taking encoding load and may enhance understanding by synthesizing information for them  \cite{fang_understanding_2022, makanyOptimisingUseNote2009}.

 In the \IntermediateAI condition, AI generates intermediate summary blocks right after each speaking turn, as shown in Figure \ref{fig:system}(a). These summaries offer a brief interpretation of lecture segments without a fully structured format, providing users with essential information that they can selectively integrate into their notes. The interval between summary blocks is around 15 seconds, where a natural sentence ends. Users can expand to check the original transcript of the summary block by clicking the block Figure \ref{fig:system}(a1).
This condition leverages the cognitive benefits of “building blocks” \cite{hou_using_2022}. By using intermediate summaries, this level of support is designed to encourage students to engage with the content actively, reducing the cognitive load associated with verbatim note-taking while still fostering their encoding process through selective integration and organization.

In the \MinimalAI condition, AI provides real-time transcripts of the lecture, divided into "transcript blocks" based on turns, as shown in Figure \ref{fig:note_transcript}(t). Users can incorporate these transcript blocks into their notes through drag-and-drop. They can also select the specific content on one transcript block and click an "insert" button to put it into the editor, shown in Figure \ref{fig:note_transcript}(t3). This condition represents low AI assistance, where the AI merely acts as a passive transcription tool without interpretation or structuring, thereby requiring the user to handle all aspects of encoding and organization. 
This design mitigates the potential cognitive overload associated with fully manual note-taking \cite{piolatCognitiveEffortNote2005b}, offering light-touch assistance that students can choose to engage with as needed.

While the Real-Time \textsc{AI-Generated Note Panel} delivers AI-generated content proactively at predefined intervals, the \textsc{Search and Synthesize Panel} empowers users to actively navigate the same AI assistance levels according to their own needs and points of interest.  In this way, it complements the proactive panel by helping participants flexibly apply the AI assistance level to match their evolving information needs throughout the lecture.
 As shown in Figure~\ref{fig:note_transcript}, the content retrieved via the search panel adopts the same granularity and abstraction as the content shown in the {AI-generated notes Panel}. 
Specifically: \begin{itemize} \item \MinimalAI: retrieves and surfaces raw transcript segments (Figure~\ref{fig:note_transcript}-t3). \item \IntermediateAI: generates summaries based on relevant transcript chunks (Figure~\ref{fig:system}-a1). \item \AutomatedAI: synthesizes structured notes from multiple segments (Figure~\ref{fig:note_transcript}-n2). \end{itemize}

We decided against a no-AI control condition because participants can fully control and opt to disregard AI-generated content in any of the three conditions; additionally, the system permits minimizing the AI panel to lessen visual distraction. Thus, users can maintain self-directed note-taking if desired.

\subsubsection{Design Considerations} \label{design consideration}

The system is intentionally designed to reflect varying degrees of AI assistance and user cognitive engagement, aiming to empirically examine the AI assistance dilemma — the tension between automation benefits and user cognitive engagement risks.

We focus on proactive AI assistance as our primary design axis to simulate real-time note-taking support scenarios. Traditionally, learners actively engage in encoding by either transcribing content verbatim or constructing more abstractive summaries. Prior research has shown that higher degrees of information processing, such as synthesizing rather than simply copying, correlate with deeper cognitive engagement during note-taking \cite{bauer2007selection, sunHowStudentsGenerate2022}. In designing our system, we intentionally align AI assistance with this natural human note-taking progression: by varying the degree of AI processing, the system can assist with this encoding process, specifically, by offering raw transcripts to support verbatim recording, concise summaries to promote selective abstraction, and structured notes to offload synthesis.
 However,  we could hypothesize that AI-generated notes reduce the effort in note-taking, which may potentially reduce the depth of cognitive engagement of students in the \textit{encoding process} of note-taking and influence comprehension. This AI assistance in the encoding stage may also shift the nature of users’ effort in the \textit{storage function} of note-taking. On the one hand, because AI generates structured notes, it may enhance review and understanding by synthesizing information for them. On the other hand, as AI generates more processed and synthesized notes, users may need to invest greater cognitive effort in reading, understanding, and critically evaluating these AI-generated outputs before using them. These are the nuanced influences introduced by varying levels of AI assistance during note-taking—both in encoding and storage—that we aim to empirically investigate in our study.

%

\begin{figure}[h!]
    \centering
    \includegraphics[width=1\linewidth]{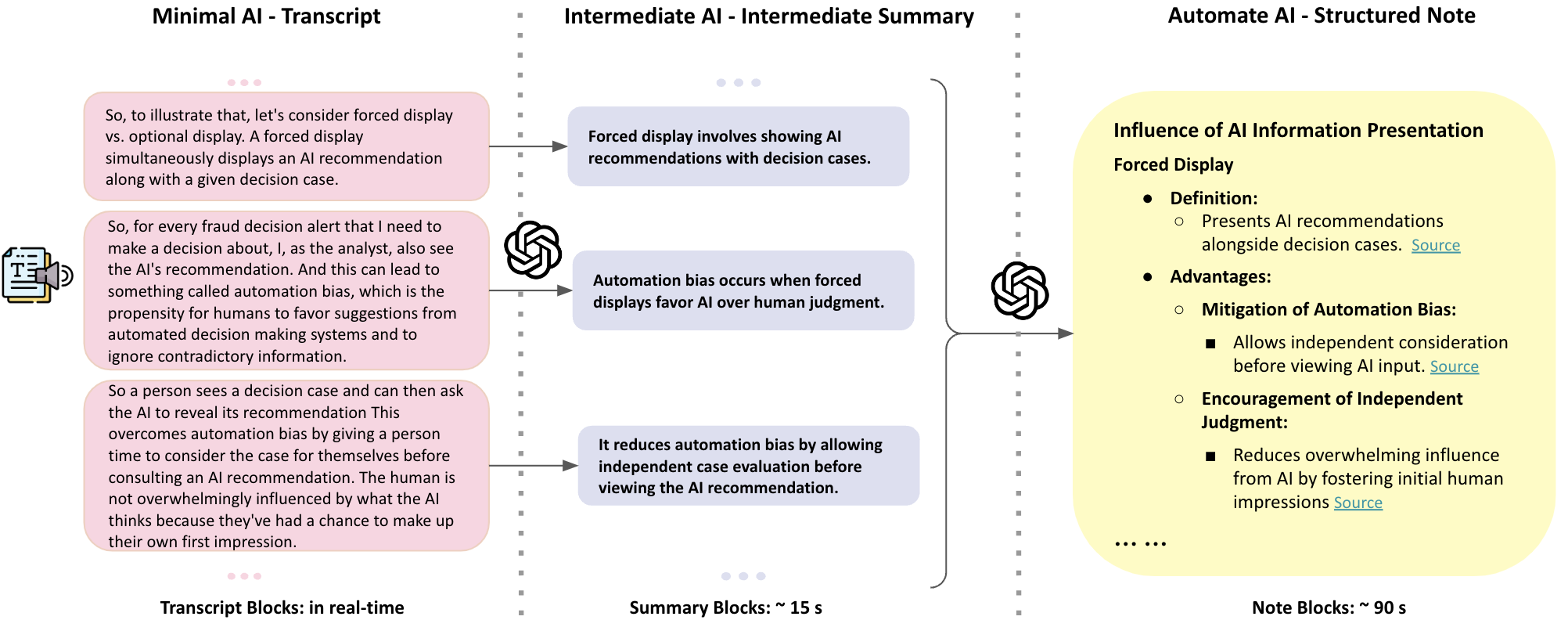}
    \caption{\textbf{The granularity of AI notes in the three conditions.} From left to right, the AI-generated results of three systems—\MinimalAI, \IntermediateAI, and \AutomatedAI—illustrate the topic of \textsc{forced display} with increasing levels of AI assistance. The \MinimalAI system presents users with raw AI-chunked transcripts, providing unfiltered information. In contrast, the \IntermediateAI system offers concise summaries of each transcript chunk, enhancing clarity and facilitating user comprehension. Finally, the \AutomatedAI system consolidates all short summaries into structured notes that include definitions and advantages of \textsc{forced display}. This organized presentation provides detailed information for each point, allowing users to reference related source transcripts for further context.}
    \label{system_comparison}
\end{figure}

We deliberately balance several critical factors in designing the varied levels of AI assistance: (1) the need to provide timely, real-time AI support during the note-taking process;
and (2) the necessity of maintaining comprehensive coverage of lecture content to ensure that AI-generated notes remain complete and reliable, regardless of their abstraction level.
To achieve this, we control the granularity and completeness of content delivery across all conditions. 
Specifically, transcript and summary blocks are consistently segmented based on natural turns—a complete sentence or a set of related short sentences—ensuring comparability in chunk size and information coverage. The \MinimalAI variant provides a real-time transcript. In the \IntermediateAI variant, AI generates summaries immediately after each turn, averaging around 15 seconds per summary block. The difference between the two is that the summary blocks are easier to read, whereas transcript blocks are more transparent since they are true representations of the conversation. The \AutomatedAI variant combines several turns into topic-based structured note blocks, with each note block between 1-2 minutes and traceable to the original transcript.  Although the abstraction level varies, all AI-generated content offers a full representation of the lecture, ensuring comparability across the three system variants, as shown in Figure \ref{system_comparison}.

\subsubsection{System Implementation}

We built NoteCopilot with a React.js frontend and Django backend. For experimental consistency, all AI-generated notes in \IntermediateAI and \AutomatedAI were pre-generated and timestamped to ensure identical content for all participants, avoiding variability in live AI output.

In \MinimalAI, Azure Speech-to-Text transcribed lectures in real-time, displaying text word-by-word. We use GPT-4o to predefine sentence-level turn segmentation (Figure~\ref{system_comparison}), controlling word count and timing. At each segmentation point, the system wrapped prior text into a transcript block and started a new one, ensuring consistent chunking across participants.
In \IntermediateAI, these segments were passed to the LLM model to produce concise summaries, which were shown immediately after each lecture turn, simulating real-time generation.
In \AutomatedAI, GPT-4o was used to group multiple transcript segments into thematic chunks and generated structured notes synthesizing these higher-level units, pre-loaded and “replayed” along the lecture timeline.

In the \textsc{Search and Synthesize Panel}, all conditions used the same prompting approach: the full transcript and user query were input to GPT-4o, with instructions tailored to the assistance level. \MinimalAI returned relevant raw transcript segments; \IntermediateAI returned synthesized summaries; \AutomatedAI returned higher-level structured notes.

To produce a well-structured, high-quality summary and notes, we spent substantial prompt engineering effort to ensure that the resulting notes appear reasonable and organized, as illustrated by the AI-generated transcript, summary, and notes shown in Figure \ref{system_comparison}. 
To ensure content accuracy, all generated summaries in \IntermediateAI and notes in \AutomatedAI were manually reviewed by the researchers to confirm their alignment with the lecture content.





\section{Data Analysis and Measurement}
We performed quantitative analyses to answer RQ1:

\begin{itemize}

    \item RQ1: How do different levels of AI assistance affect learning outcomes, cognitive load, note-taking behaviors, and usability?
\end{itemize}

We then performed qualitative analysis on our interview data to answer RQ2:
\begin{itemize}
     \item RQ2: Why do these effects occur? How do students perceive the usefulness of AI, and how do they prefer to interact with different AI assistance levels?
\end{itemize}

\subsection{Quantitative Outcome Measures}


\subsubsection{Measurement of Understanding}
To assess users' understanding of the lecture, we graded students' two rounds of post-test answers, i.e., immediately after the lecture, and after the participants revised their answers based on the notes. The first round, completed without reference to notes, measured students' real-time understanding of the lecture content, referred to as the \textsc{Post-Test Score}. The second round allowed students to revise their initial answers using their notes, called the \textsc{Revision Score}. This score represents the understanding achieved after reviewing and refining their responses with their notes.

We developed a rubric to evaluate answers to open-ended questions (OEQ). One author initially created the rubric, with each question worth up to 4 points. Two authors then independently graded a sample (25\%) of the students' answers using this rubric, blind to the experimental condition. The initial inter-coder reliability was 86.7\%. After discussion and consensus on the grading standards, they divided the remaining answers for individual grading. 
The rubric and example students' answers are appended in the supplementary materials \footnotemark[\getrefnumber{resource}].

\subsubsection{Measurement of Cognitive load}
We assessed cognitive load by employing questions tailored to evaluate overall mental effort \cite{paas1993efficiency}, and we adapted scales from earlier studies \cite{klepsch2017development, leppink2013development} to measure intrinsic (IL), extraneous (EL), and germane (GL) cognitive loads. We opted not to use a general task load measurement, such as NASA TLX, and instead chose a scale specifically designed to measure the three types of cognitive load. This approach allows us to examine how each type of cognitive load is influenced by the three conditions, providing insights into how varying levels of AI assistance can enhance beneficial aspects of cognitive load, like germane load, while mitigating the more negative aspects, such as extraneous load.

\paragraph{Overall Mental Effort:}
Please rate your overall mental effort (from 1 - 9). Scale 1 = very, very low mental effort, and Scale 9 = very, very high mental effort \cite{paas1993efficiency}
\paragraph{Intrinsic Load:}
This type of cognitive load refers to the cognitive effort required to understand the inherent complexity of the learning content \cite{klepsch2017development}. It depends on the difficulty of the material itself, independent of the instructional methods.
Questions include:
\begin{itemize}
    \item Q1: "The content explained in this lecture is very complex."
    \item Q2: "This lecture contains many terms and concepts that are unfamiliar to me."
\end{itemize}
The score ranged from 1 (strongly disagree) to 7 (strongly agree).

\paragraph{Extraneous Load:}
Extraneous load is the cognitive burden imposed by the instructional methods or tasks, which arise from unnecessary or confusing elements in the learning environment \cite{klepsch2017development}.  Questions related to extraneous load include:
\begin{itemize}
    \item Q3: "The instructor failed to explain concepts clearly."
    \item Q4: "Manual note-taking required a lot of my effort."
    \item Q5: "Interacting with the system (e.g., using AI-generated notes) required significant effort."
    \item Q5: "Creating notes with AI assistance also required significant effort."
    \item Q6: "Reading AI-generated notes required significant effort."
\end{itemize}
The score ranged from 1 (strongly disagree) to 7 (strongly agree).

\paragraph{Germane Load:}
Germane load is the cognitive effort devoted to processes that directly support learning, understanding, and knowledge construction \cite{klepsch2017development}. This is a type of beneficial load that helps deepen comprehension and consolidate knowledge. Questions include:
\begin{itemize}
    \item Q7: "I was fully engaged in this lecture."
    \item Q8: "Note-taking with AI assistance enhanced my understanding of the lecture."
    \item Q9: "Reading AI-generated notes enhanced my understanding of the lecture."
\end{itemize}
The score ranged from 1 (strongly disagree) to 7 (strongly agree).

\subsubsection{Note-taking Behaviors and Outcomes}

Throughout the experiment, we systematically logged users' interactions with the interface. This included actions on the \textsc{Real-time AI-generated Note Panel},  \textsc{Search and Synthesize Panel}, and \textsc{text-editor}.  We also recorded the content created in the Text Editor, distinguishing between notes generated by AI and those created by users.  By analyzing this log data, we aim to gain a detailed understanding of how participants utilized the AI assistance in their note-taking process and how this influenced their sense-making process. 

\paragraph{Note-Taking Behaviors} 
The study tracked several note-taking behaviors, which included:
\begin{itemize}
    \item \textsc{Clicking on the Text Editor}: Frequency of checking the content in the text editor
    \item \textsc{Typing in the Editor}: Frequency of typing actions in the text editor
    \item \textsc{Dropping AI-Generated Content}: Inserting AI-generated content into the editor through  drag and drop or the 'insert' button
    \item \textsc{Query Search}: Searching via direct queries in the \textsc{Search and Synthesize Panel}
    \item \textsc{In-Context Search for Previous Content}: Searching for previously discussed content around the selected texts in the text editor
    \item \textsc{In-Context Search for Upcoming Content}: Searching for future content around the selected texts in the text editor
\end{itemize}

\paragraph{Note Outcome Measures} 
The outcome of the note-taking process was measured based on \textsc{note quantity}:
\begin{itemize}
    \item \textsc{Total Words in the Editor}: The sum of all words, including both AI-generated content used by participants and manually typed content.
    \item \textsc{Manually Typed Words}: The count of words manually typed by participants in the editor.
\end{itemize}

This study only measured the \textsc{quantity of notes} and did not assess \textsc{note quality}. We acknowledge this as a limitation of our research and will address it in the discussion.

\subsubsection{Usability}
Usability was measured through a series of subjective usability questionnaire items, including scores for system ease of use, satisfaction, future usage intention, etc.


\subsection{Interview Analysis}
To understand students' perceptions of the usefulness and limitations of the three levels of AI assistance in note-taking, we conducted a thematic analysis of interview transcripts \cite{braun2012thematic}. Initially, two researchers independently reviewed, commented on, and coded the transcripts. They developed 203 initial codes in this phase. They then discussed their findings to reach a consensus, after which one researcher refined the codes and identified key themes across the transcripts. All disagreements between coders were resolved through discussion.  We uncovered 8 themes related to the benefits and limitations of each condition, 3 themes on how students used different AI assistance levels in note-taking, and 7 themes covering students' perceptions of agency, intention expression, needs, and trustworthiness. From these themes, we derived 5 high-level findings, which focused on AI's impact on students' learning, when and how students expressed their intentions toward the AI, and students' perceptions of agency and cognitive load.

\section{Finding}

\subsection{RQ1: How do different levels of AI assistance affect comprehension, cognitive load, note quality, and usability?}

In RQ1, we evaluated how the different levels of AI assistance (\AutomatedAI, \IntermediateAI, \MinimalAI) influence students' learning and understanding, cognitive load, and usability. 

\subsubsection{Understanding Levels}
To evaluate the impact of different levels of AI assistance (\AutomatedAI, \IntermediateAI, \MinimalAI) on students’ understanding of the lecture,  we conducted mixed-effects linear regression analyses on both post-test and revision scores. In these models, we included \textsc{Condition} as a fixed factor, controlled for \textsc{Video} as a covariate, and incorporated interaction terms between \textsc{Condition} and \textsc{Video}. Random intercepts for \textsc{Individual Participants} were added to account for within-subject variability.

\begin{table}[ht] \centering \begin{tabular}{lcc} \toprule \textbf{Coefficient (Std. Err)} & \textbf{Post-Test Score} & \textbf{Revision Score} \\
Intercept (\AutomatedAI) & $9.550$ (0.980) & $12.033$ (0.922) \\ Condition: \IntermediateAI & \textbf{$4.172$ (1.315) **} & \textbf{$2.848$ (1.190) *} \\
Condition: \MinimalAI & $2.702$ (1.339) * & $0.367$ (1.216) \\
Video: Video 2 & $2.326$ (1.338) & $0.565$ (1.216) \\ 
Video: Video 3 & $-0.298$ (1.368) & $-0.278$ (1.190) \\
Condition: \IntermediateAI × Video 2 & $-1.146$ (1.982) & $-0.382$ (1.806) \\ 
Condition: \MinimalAI × Video 2 & $-2.197$ (1.974) & $2.528$ (1.813) \\
Condition: \IntermediateAI × Video 3 & $-0.136$ (1.933) & $1.339$ (1.779) \\ 
Condition: \MinimalAI × Video 3 & $-0.993$ (1.997) & $-0.538$ (1.805) \\
Group Var & $2.216$ (0.603) & $4.096$ (0.894) \\
\bottomrule
\end{tabular} 
\caption{ Mixed-effects linear regression model results for \textsc{Post-Test Scores} and \textsc{Revision Scores}, examining the fixed effects of \textsc{Condition} and \textsc{Video} with random intercepts for \textsc{User}.  The model controlled for the influence of \textsc{Video} and included interaction terms between \textsc{Condition} and \textsc{Video}.   Results indicate that the \IntermediateAI condition led to significantly higher \textsc{Post-Test Scores} and \textsc{Revision Scores} compared to \AutomatedAI, with \MinimalAI showing a higher \textsc{Post-Test Score} than \AutomatedAI, and no significant interactions between \textsc{Condition} and \textsc{Video}. Significance codes: $^{***}$ p<0.001, $^{**}$ p<0.01, $^{*}$ p<0.05. } 
\label{score}
\end{table}

The analysis of \textsc{Post-Test Scores} revealed a significant main effect of \IntermediateAI on learning outcomes. Specifically, students in the \IntermediateAI condition (Mean = 13.88, SD = 2.34) scored significantly higher than those in the \AutomatedAI condition (Mean = 10.28, SD = 3.75), with a coefficient of 4.17 (
$p=0.002$). The \MinimalAI condition also showed a significant improvement over \AutomatedAI, with a coefficient of 2.70 (
$p=0.044$). Post-hoc Tukey HSD analysis further confirmed these differences, showing that \IntermediateAI outperformed \AutomatedAI with a mean difference of 3.61 (
$p<0.001$), and significantly outperformed \MinimalAI with a mean difference of 2.03 (
$p=0.034$).

The analysis of \textsc{Revision Scores} revealed a significant main effect of \IntermediateAI on post-review learning outcomes. Specifically, students in the \IntermediateAI condition (Mean = 15.29, SD = 2.99) scored significantly higher than those in the \AutomatedAI condition (Mean = 12.12, SD = 2.76), with a coefficient of 2.85 ($p=0.017$). The \MinimalAI condition showed no significant improvement over \AutomatedAI ($Coef = 0.37, p=0.763$).
Post-hoc Tukey HSD analysis supported these findings, showing that \IntermediateAI outperformed \AutomatedAI \ with a mean difference of 3.17 ($p=0.0003$), and significantly surpassed \MinimalAI with a mean difference of 2.00 ($p=0.031$).

These findings reveal that \IntermediateAI better supports real-time learning (\textsc{Post-Test Scores})  compared to \AutomatedAI. However, contrary to our initial expectations, \IntermediateAI also outperforms \AutomatedAI in post-review tasks (\textsc{Revision Scores}). Interestingly, despite the opportunity for students to review with notes \AutomatedAI, this condition did not yield better learning outcomes.

No significant difference between \textsc{Video} and the learning outcome was found. Hence, we argue that the difficulty levels of the videos are comparable, so we excluded video differences from further analyses in this paper.

\subsubsection{Cognitive Load}

We analyzed cognitive load across different AI-assisted note-taking conditions (\AutomatedAI, \IntermediateAI, and \MinimalAI) using repeated-measures ANOVA, focusing on dimensions of mental effort, intrinsic, extraneous, and germane cognitive load.

\begin{table}[ht]
\small
\centering
\begin{tabular}{p{5.5cm}|p{1.9cm}|p{2.3cm}|p{1.6cm}|c}
\hline
\textbf{Cognitive Load Question} & \textbf{\AutomatedAI} & \textbf{\IntermediateAI} & \textbf{\MinimalAI} & \textbf{F (Pr > F)}  \\ \hline
\textsc{Overall}: Please evaluate your mental effort in the task          & 4.60 (2.08)           & 4.53 (1.74)              & 5.17 (1.39)         & 1.6716              \\ \hline
\textsc{Intrinsic}: The learning content covered in this lecture was very complex & 3.03 (1.59)           & 2.53 (1.17)              & 3.00 (1.14)         & 1.9618                  \\ \hline
\textsc{Intrinsic}: The lecture included many terms and concepts that were unfamiliar to me & 3.13 (1.61)           & 3.13 (1.63)              & 3.57 (1.65)         & 1.1086                \\ \hline
\textsc{Extraneous}: The instructor did not explain the concepts clearly in this lecture & 2.40 (1.28)           & 2.30 (1.21)              & 2.10 (1.03)         & 0.7346                  \\ \hline
\textsc{Extraneous}: It requires a lot of effort for me to write notes by myself. & 2.50 (1.68)           & 2.90 (1.69)              & 3.77 (1.43)         & \textbf{4.3899*  }                \\ \hline
\textsc{Extraneous}: Engaging with the system (e.g., using the AI-generated notes) took a lot of effort from me. & 2.87 (1.50)           & 3.17 (1.18)              & 3.90 (1.65)         & \textbf{4.3497*   }              \\ \hline
\textsc{Extraneous}: It requires a lot of effort for me to create notes with the AI's help. & 2.87 (1.53)           & 2.83 (1.23)              & 3.93 (1.76)         & \textbf{4.7137* }                   \\ \hline
\textsc{Extraneous}: It requires a lot of effort for me to read the AI-generated notes. & 2.93 (1.66)           & 3.07 (1.70)              & 4.57 (1.61)         & \textbf{9.2956*** }                   \\ \hline
\textsc{Germane}: I am fully engaged during the lecture. & 4.33 (1.53)           & 4.77 (1.52)              & 4.50 (1.61)         & 1.0331                    \\ \hline
\textsc{Germane}: Taking notes (with the assistance of AI) enhanced my understanding of the lecture. & 4.93 (1.11)           & 5.23 (1.04)              & 4.43 (1.52)         & \textbf{4.1359* }                  \\ \hline
\textsc{Germane}: Reading the AI-generated notes enhanced my understanding of the lecture. & 5.23 (1.10)           & 4.83 (1.18)              & 4.23 (1.50)         & \textbf{4.4918* }                  \\ \hline
\end{tabular}
\caption{Repeated Measures ANOVA Results for Cognitive Load Questions across Conditions with Mean (SD) for Each Condition}
\label{table:cognitive_load_anova_means}
\end{table}

\paragraph{Overall Mental Effort}
The repeated-measures ANOVA for overall mental effort showed no significant main effect of condition (\(F(2, 58) = 1.67\), \(p = 0.197\)), indicating similar levels of mental effort across conditions. 

\paragraph{Intrinsic Cognitive Load}
For intrinsic cognitive load, measured by perceived complexity and unfamiliarity with lecture content, no significant main effect of condition was observed. These findings suggest that intrinsic load was not significantly impacted by the AI assistance levels, aligning with cognitive load theory, which posits that intrinsic load is driven primarily by the inherent complexity of the content rather than instructional conditions.

\paragraph{Extraneous Cognitive Load}
Significant effects were observed for extraneous cognitive load across conditions, particularly in manual effort, AI engagement, and reading AI notes.
For the effort in manual note-taking, \MinimalAI (Mean = 3.77, SD = 1.43) required significantly more effort than \AutomatedAI (Mean = 2.50, SD = 1.68, \(F = 4.39\), \(p = 0.017\)), indicating that manual note-taking without AI support increases extraneous load.
Similarly, for engaging with the system, \MinimalAI (Mean = 3.90, SD = 1.65) required significantly more engagement effort than \AutomatedAI (Mean = 2.87, SD = 1.50, \(F = 4.35\), \(p = 0.017\)).
For reading AI-generated notes,  \MinimalAI (Mean = 4.57, SD = 1.61) required significantly higher reading effort than both \IntermediateAI (Mean = 3.07, SD = 1.70, 
$p=0.002$) and \AutomatedAI (Mean = 2.93, SD = 1.66, 
$p=0.0007$) \((F=9.30,  p<0.001)\).

\paragraph{Germane Cognitive Load}
For germane cognitive load, significant effects were observed for both questions related to the interaction with AI. The question, \textit{"Taking notes (with the assistance of AI) enhanced my understanding of the lecture,"} showed a significant effect of condition (\(F = 4.14\), \(p = 0.021\)), with \IntermediateAI (Mean = 5.23, SD = 1.04) significantly outperforming \MinimalAI (Mean = 4.43, SD = 1.52, \(p = 0.038\)). Similarly, \textit{"Reading the AI-generated notes enhanced my understanding of the lecture,"} yielded a significant effect (\(F = 4.49\), \(p = 0.015\)), with \AutomatedAI(Mean = 5.23, SD = 1.10) significantly higher than \MinimalAI (Mean = 4.23, SD = 1.50, \(p = 0.009\)). 

In summary, \AutomatedAI is optimal for lowering extraneous load, while \IntermediateAI \\ supports germane cognitive load by using AI-generated Notes. Conversely, \MinimalAI demands greater manual effort, showing higher extraneous load and challenging cognitive resources.

\subsubsection{Note-taking behavior}
We analyzed the note outcomes and note-taking behaviors across different AI-assisted note-taking conditions using repeated-measures ANOVA. 

\begin{figure}[!h]
    \centering
    \includegraphics[width=.95\linewidth]{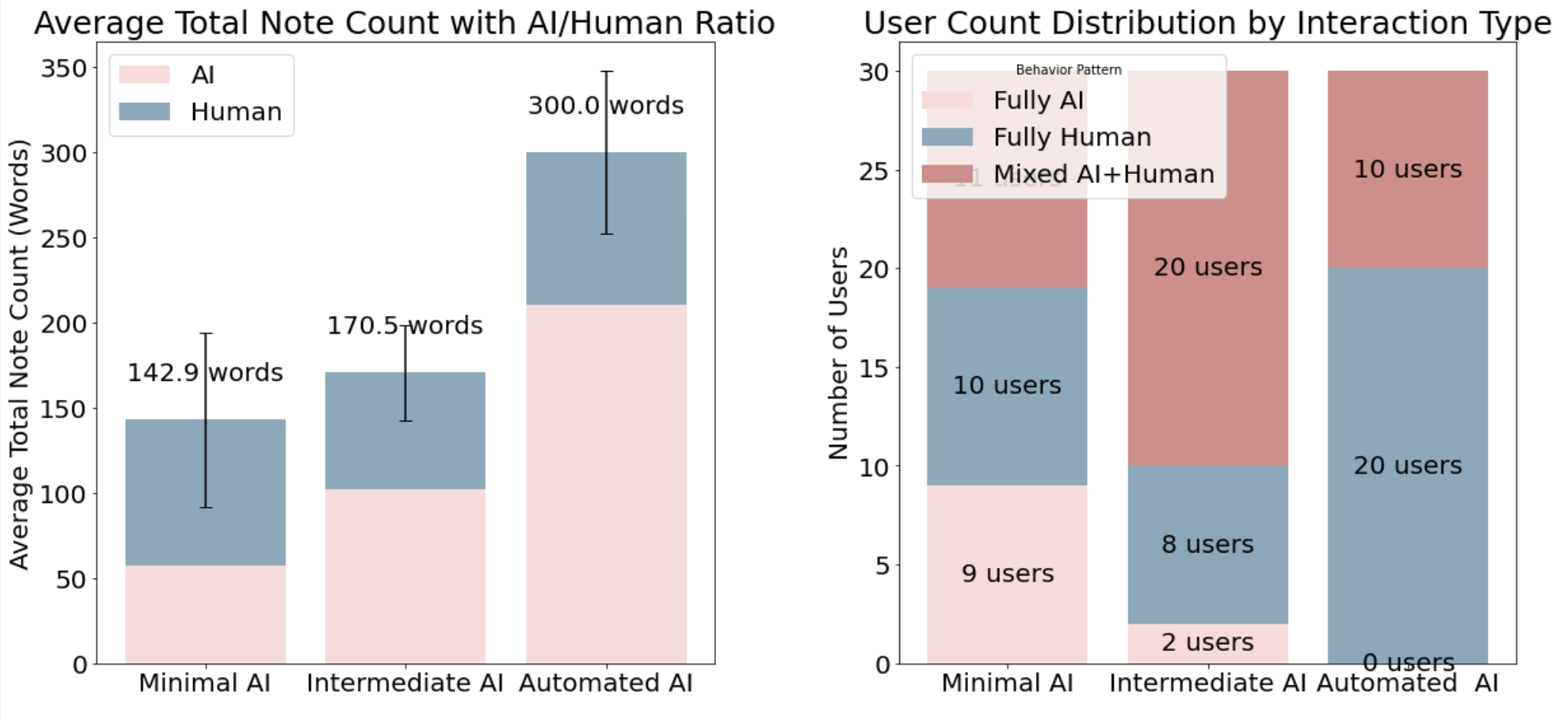}
    \caption{\textbf{Comparison of note-taking behaviors across conditions with varying levels of AI assistance.} The left chart shows the average total note count for each condition, divided into AI-generated content and manually typed content, with error bars indicating standard deviation. On the right, a stacked bar chart illustrates the distribution of users by interaction type (fully AI-generated, mixed AI-human, or fully manual) across conditions, with labels indicating the number of users per type.  
    }
    \label{fig:Note Outcome}
\end{figure}

There are notable differences in average total note counts across different AI conditions, as shown in Figure \ref{fig:Note Outcome}. The \AutomatedAI condition yielded a significantly higher total note count (Mean=300.68,  SD=160.52) compared to both the \IntermediateAI condition (Mean=170.46
SD=70.21) and the \MinimalAI condition (Mean=143.69, SD=86.93). ANOVA results confirmed a significant effect of AI condition on note volume
($F=16.66,  p<0.001$). 

To further understand these differences, we classified users into three general note-taking patterns based on their approach to AI assistance: Fully AI Mode, where users only used AI-generated notes without adding their own; Mixed AI+Human, where users combined AI notes with their manual entries; and Fully Manual, where users relied solely on self-generated notes. The distribution of these strategies across conditions reveals distinct behavioral tendencies. In the \IntermediateAI condition, over half of the users adopted the Mixed Mode, integrating AI-generated notes with their manual inputs. A clear divergence emerged in the \AutomatedAI condition: 10 users chose to take their notes manually, while 8 relied solely on AI.  In the \MinimalAI condition, users predominantly engaged in fully manual note-taking, with 19 users relying exclusively on human-generated notes and 10 employing a mixed approach.  



Further analysis of interaction behaviors in each condition offered insights into users’ note-taking engagement. Users in the \IntermediateAI condition frequently interacted with AI-generated notes, as indicated by the highest click counts in the text editor (Mean = 37.41, SD = 24.86) compared to the \MinimalAI condition (Mean = 28.14, SD = 18.47) ($p=0.049$). Typing frequency was also higher in the \IntermediateAI condition (Mean = 37.59, SD = 13.33), significantly surpassing \MinimalAI ($p=0.008$). AI content “drops” (when AI-generated text was selectively added to the editor) were also most frequent in the \IntermediateAI condition (Mean = 8.03, SD = 7.51), demonstrating a more active and curated approach than \AutomatedAI (
$p<0.001$) and the \MinimalAI condition ($p<0.001$). Interaction with initiating a query and constantly listening to the upcoming content was notably higher in the \IntermediateAI condition (Mean = 7.38, SD = 30.44), compared to that in \AutomatedAI (Mean = 1.93, SD=4.96) ($p<0.001$).

\begin{table}[!h]
\centering
\small
\begin{tabular}{lcccc}
\hline
Behavior & \MinimalAI & \IntermediateAI & \AutomatedAI & F (Pr>F) \\
\hline
\textbf{Manual type words} & 110 (83.58) & 86.58 (56.01) & 94.46 (61. 62) & F = 1.42 \\
\textbf{Click Text Editor} & 28.14 (18.47) & 37.41 (24.86) & 32.07 (24.46) & F = 3.18 * \\
\textbf{Times for Typing words} & 30.21 (15.59) & 37.59 (13.33)  & 28.97 (12.12) & F = 0.86 \\
\textbf{Drop AI Content} & 1.14 (1.94) & 8.03 (7.51)  & 3.00 (2.93) & F = 4.58 * \\
\textbf{Search through query} & 0.66 (1.26) & 0.55 (1.59) & 0.52 (1.30) & F = 0.66 \\
\textbf{In-context search} & 0.24 (0.64) & 0.14 (0.44) & 0.00 (0.00) & F = 1.42  \\
\textbf{Listen upcoming content} & N/A & 7.38 (30.44)  & 1.93 (4.96) &  N/A \\
\hline
\end{tabular}
\caption{Summary of Note-Taking Behaviors During the Lecture Stage Across Conditions, with ANOVA Results for Each Behavior}
\end{table}

Are there more AI drop behaviors in \IntermediateAI than in \AutomatedAI due to different frequencies of AI content generation? To further unpack the differences in user engagement with AI-generated content across conditions, we conducted an additional analysis focusing on AI content use behavior. Specifically, we examined (1) the total number of knowledge units dropped into the editor, and (2) the number and proportion of knowledge units that were subsequently modified by users. Here, we define a "knowledge unit" as a semantic segment of AI-generated content: in \MinimalAI, this corresponds to a transcript block; in \IntermediateAI, a summary block; and in \AutomatedAI, a source-linked knowledge point rather than the full blocks, as illustrated in Figure~\ref{system_comparison}.  We further analyzed the extent to which users modified these AI-generated knowledge units after dropping them into the editor. Modification counts were computed by comparing the AI-generated content dropped into the editor with the final version of the user’s notes.

\begin{table}[!h]
\centering
\small
\begin{tabular}{p{4.5cm}cccc}
\hline
Metric & \MinimalAI & \IntermediateAI & \AutomatedAI & F(Pr>F)  \\
\hline
Total AI content drops (normalized by knowledge units) & 1.14 (1.94) & 8.03 (7.51) & 10.03 (4.89) & F=57.19 *** \\
Modified knowledge units (count) & 0.30 (0.46) & 3.53 (2.04) & 0.76 (0.93) & F=66.37 *** \\
Modified knowledge units (\% of dropped) & 26.3\% & 40.0\% & 8.0\%  & F=23.60 ***\\
\hline
\end{tabular}
\caption{Normalized analysis of AI content use across conditions. To account for the varying frequency and granularity of AI-generated content, we normalize the total AI content drops by the number of knowledge units. We further report the number and proportion of knowledge units that were subsequently modified by users, estimated by comparing the dropped AI content with the final editor content.}

\label{tab:normalized_ai_drops}
\end{table}

As shown in Table~\ref{tab:normalized_ai_drops}, we found that the normalized number of AI knowledge units dropped into the editor was comparable between the \IntermediateAI (Mean = 8.03, SD = 7.51) and \AutomatedAI (Mean = 9.41, SD = 4.89) conditions ($p = 0.07 > 0.05$). However, users in the \IntermediateAI condition demonstrated substantially higher modification behaviors. On average, they modified 3.53 units compared to 0.77 in \AutomatedAI ( $p < 0.001$), and 0.30 in \MinimalAI ($p < 0.001$).
These findings show some evidence that \IntermediateAI encouraged more active refinement of AI-generated content, whereas \AutomatedAI led users to more passively accept AI outputs with minimal modification. We acknowledge the limitations of our current analysis approach. While we normalized the number of AI knowledge units and captured the count of modified units, our method only reflects the end-state differences in user edits. It does not capture the dynamic editing processes that may occur during note-taking, such as additions, reorganization, or deletions of AI-generated content. 

\subsubsection{Usability}

We used repeated measures ANOVA to assess usability across different levels of AI assistance, with post-hoc Tukey HSD tests to identify specific pairwise differences between conditions.

\begin{figure}[!h]
    \centering
    \includegraphics[width=1\linewidth]{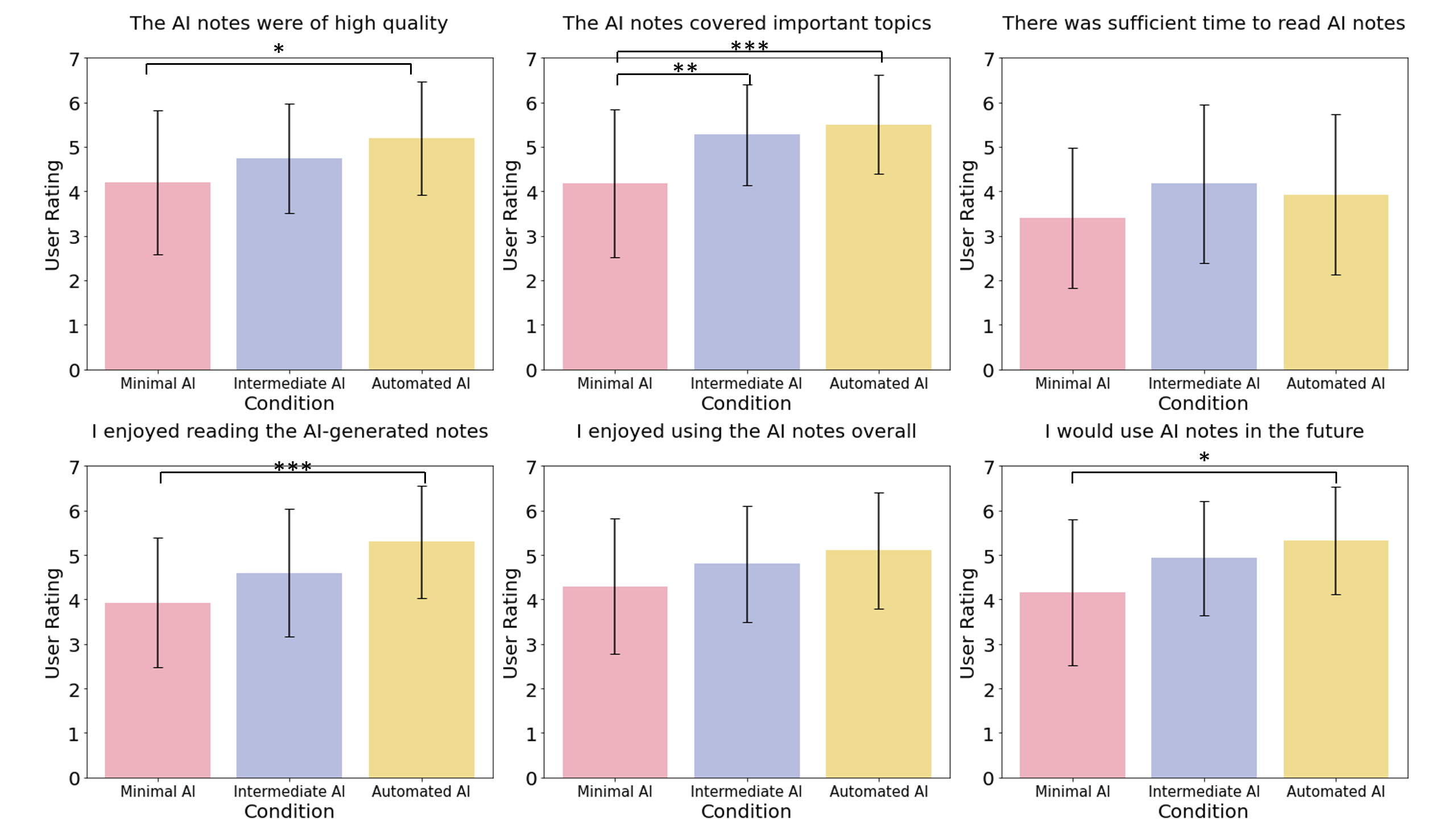}
    \caption{Usability Rating. This figure presents the mean usability ratings across three conditions—\MinimalAI, \IntermediateAI, and \AutomatedAI —for six questions assessing different aspects of the AI-generated notes. Tukey's HSD pairwise comparisons were conducted after the repeated-measure ANOVA. .p-values are marked as follows: *p < 0.05, **p < 0.01, ***p<0.001. The non-significant difference was annotated as NS.}
    \label{fig:enter-label}
\end{figure}

In the \AutomatedAI\  condition, participants rated the perceived quality of AI-generated notes significantly higher (Mean = 5.20, SD = 1.27) compared to the \MinimalAI\  condition (Mean = 4.20, SD = 1.61), with \(F = 3.95, p = 0.025\). Additionally, for coverage of important topics, \AutomatedAI\ (Mean = 5.50, SD = 1.11) was rated significantly higher than \MinimalAI\ (Mean = 4.17, SD = 1.66) (\(F = 9.50, p < 0.001\)). These results suggest that users found \AutomatedAI's output of notes more effective in capturing key points.
Enjoyment in reading AI-generated notes was also substantially higher in the \AutomatedAI\ condition (Mean = 5.30, SD = 1.26) compared to \MinimalAI\ (Mean = 3.93, SD = 1.46) (\(F = 8.18, p < 0.001\)). Furthermore, the intention for future use was notably higher in the \AutomatedAI\ condition (Mean = 5.33, SD = 1.21) compared to \MinimalAI\ (Mean = 4.17, SD = 1.64) (\(F = 5.62, p = 0.006\)).

In the \IntermediateAI\ condition, participants rated coverage of important topics significantly higher (Mean = 5.27, SD = 1.14) than in \MinimalAI\ (Mean = 4.16, SD = 1.66) (\(p\) = 0.005).  Besides, \MinimalAI\ (Transcript) condition shows a higher preference for taking manual notes (Mean = 3.80, SD = 1.56) than those in the \AutomatedAI\ (Auto-Note) condition (Mean = 2.63, SD = 1.45), with \(p = 0.01\), as shown by post-hoc Tukey HSD tests. This suggests that users prefer manual note-taking more when AI is seen as unusable. We didn't see other difference in the usability rating between \IntermediateAI and \MinimalAI or between \IntermediateAI and \AutomatedAI

In summary, \AutomatedAI\ demonstrates optimal usability, significantly enhancing perceived quality, topic relevance, and enjoyment in reading AI-generated notes. \IntermediateAI\ offers a balanced usability experience, showing improved coverage of important topics over \MinimalAI, though it does not reach the high preference and engagement levels of \AutomatedAI. Conversely, \MinimalAI\ has lower usability ratings, with a higher preference for manual note-taking, suggesting that increased manual effort may detract from the perceived utility and enjoyment of AI-assisted notes.

\subsection{RQ2. How Do Students Perceive AI's Helpfulness in Note-Taking? }
For RQ2, we conducted a qualitative analysis to explore how students perceive the helpfulness of AI in the note-taking process, focusing on when specific AI conditions were considered useful or not and why certain conditions were ultimately less utilized despite initial perceptions.

\subsubsection{\IntermediateAI enhanced students' learning by fostering active engagement with synchronous and digestible information.}

Students found the summaries valuable for two main reasons. First, the summaries scaffold their cognitive engagement. Beyond serving as a "constant flow" to help them stay focused on lecture content in real-time (P13, P15), they felt that interacting with summary blocks actively deepened their understanding. This involved \textit{"quickly glance over the summaries to check if key points were there"}(P5), \textit{“which summaries are relevant and which ones they need”}(P3), and \textit{"trying to put it under my own structure"}(P1). As P3 said, \textit{"I'm just gonna drag the summaries of the last few things that you talked about onto here, and then I can kind of piece that together, and then that catches me up at least a little bit more than I normally would, where I would just miss that whole chunk of information."}  
As a result, students reported substantial learning gains in the \IntermediateAI condition (P1, P13, P15).


The concise, digestible format of the \IntermediateAI summaries makes engagement possible in real-time lectures. P2 remarked that the summaries helped them focus on the lecture by breaking down information into \textit{"digestible chunks that were easy to put into the notes."} The granularity of the summaries matched many students’ natural note-taking styles, allowing seamless integration into their notes. As students highlighted, the summaries consisted of short sentences they would typically write, enabling them to \textit{"catch up by just dragging and dropping the summaries"} (P15) and do so \textit{"without needing to add further details"} (P20).

\subsubsection{The \AutomatedAI was perceived as beneficial for its well-structured content, but it also led to feelings of disengagement and inflexibility.}

Many students perceived \AutomatedAI as valuable during lectures due to their clear and logical structure. For instance, P3 and P13 noted that the structured format \textit{"organized and contextualized the information better than other conditions."} 
The structure allowed P4 to \textit{"work at their own pace and reference it later."} Students also highlighted the value of structured notes for post-lecture review and information retrieval. For example, P12 used the notes to confirm their understanding of concepts, while P15 used them to recall connections between topics. Due to the clear structure, students \textit{"knew exactly where they could find a piece of information"} (P11) and could \textit{"easily find a relevant bullet point by scrolling through the notes"} (P4). 

However, structured content and the seemingly complete nature of notes are perceived to lead to less cognitive engagement.
P3 admitted to \textit{"zoning out"} at times, despite recognizing that the topic required their full attention. Some students felt they retained only a general impression of the course content due to over-relying on the notes, 
as noted by P5,\textit{"I was struggling with answering the questions because I thought notes would help, but then I couldn't remember where it is. I felt like some sort of disconnection."} 
Lastly, the granularity of the notes presented challenges; P2 found it difficult to work with, commenting, \textit{"everything was displayed in a giant chunk,"} which hindered their ability to integrate information into their own notes.

\subsubsection{\MinimalAI has a higher cognitive load but also retains the highest level of trust.}
Many students experienced the highest cognitive load during the \MinimalAI condition. Firstly, the verbal text made it challenging to parse and transfer information (P13). Secondly, students felt distracted by the constant need to switch between interpreting the transcript, editing, and listening to the lecture. For example, P12 found it \textit{"time-consuming to decipher the transcript while trying to keep up with the lecture"}. 

However, in this condition, the AI was often seen as a \textit{"shadow companion for real-time understanding,"} helping students capture specific terminology or follow along with the lecture content as closely as possible. Since the AI only performed basic grammatical and punctuation corrections, students felt that it \textit{"reflected the lecture content as closely as possible"} (P29) and \textit{"did not add the cognitive burden of learning additional AI-generated content"} (P17). Compared to the other two conditions, although the cognitive load was highest here, the AI rarely introduced misunderstandings 
(P9).
In contrast, the other two conditions (\AutomatedAI, \IntermediateAI) occasionally led to uncertainty. Users noted that they felt \textit{"forced to adopt a new logic to interpret the AI content, which was impossible when I hadn’t fully understood the lecture in the first place"} (P19).

\subsubsection{Students valued note-taking as a learning opportunity and preferred maintaining control over their own notes.}
Students value the agency they have over their notes. Most students viewed AI assistance as a \textit{"backup"} (P20) or \textit{"augmentation"} (P8) rather than a replacement, using it selectively for tasks like verifying information (P5), catching up on missed points (P10), and enhancing note completeness (P20). 
Although students found AI support helpful, they emphasized the importance of manual note-taking as a critical learning opportunity. Writing original notes allowed them to \textit{"memorize better"} (P11, P12), \textit{"avoid distraction and keep up with the lecture"} (P13), \textit{"engage actively in the lecture instead of passively copying information"} (P15), and \textit{"gain a deeper understanding of concepts"} (P20). 
In the \AutomatedAI condition, even with AI-generated structured notes, some students (10/30) still preferred to maintain their own notes rather than use the AI-generated content and found it \textit{"easier to retrieve information"} when notes followed their own structure (P4). 

\subsubsection{Students actively seek real-time clarification and AI content to align with their goals.}

Participants actively seek AI assistance through searching to help \textit{"capture missed points when the lecture went fast"} (P6) and to \textit{"efficiently access specific information"} when the sheer volume of notes felt overwhelming (P4). After the lecture, students sought more information from the AI to clarify concepts they didn’t fully understand or recall. For instance, P1 noted, \textit{"I think the notes didn’t go deep into its pros and cons, so I asked it to generate more notes on this."}.Additionally, students valued the ability to perform \textit{"in-contextual searches"}; as P28 noted, \textit{"listening to the future makes sense since I can selectively get some results related to that topic constantly, which saves my time reading."}

Interestingly, some students used the search feature not just to fill gaps but also to \textit{"verify the accuracy of AI-generated notes"}—even for concepts they felt they already understood. In the \AutomatedAI condition, P5 searched for a specific concept to assess the AI’s accuracy, but when the result didn’t meet expectations, it reduced their trust in the AI. 
While the search feature enabled active engagement, 
students felt uncertain when searching for notes on future topics. P1 described this uncertainty: \textit{"I'm not sure whether the teacher will talk about it."}

\subsubsection{Users anticipate more contextual and intent-aware AI assistance
}\label{finding - intent}

Students also reported that it is important for AI assistance to align with their note-taking preferences and the nature of the task. For instance, some students felt that \IntermediateAI would be particularly helpful in conceptually dense lectures, where they wanted to stay engaged and needed scaffolding to keep up with the content while still thinking actively, like P4 noted, \textit{"If I want to understand something, I like having the summary—it gives me something to build from."}
When the goal was to retain detailed information for later study, some students preferred more complete outputs like those in the \AutomatedAI condition.  “\textit{For a review-heavy class,”} P8 explained, \textit{“I just want something I can look back on—structured notes are better for that.”}

In addition to this contextual variation, several participants expressed a desire for AI support that could better adapt to their evolving needs. They imagined that the system could gradually learn their preferences based on how they reorganized or edited the AI-generated content. As P19 shared, \textit{"If I’m always rewriting its definition style, it should know I like it shorter and with examples."} 
Students hoped that the system could infer their focus or confusion based on interaction patterns. As P23 noted, \textit{"If I spend a while hovering or editing a part, maybe it could suggest related points."}





\section{Discussion}

Our research centers around the questions of whether high levels of AI assistance harm students' cognitive engagement and comprehension and whether providing intermediate outputs can mitigate some of these negative effects. Our findings indicate that while \AutomatedAI offers the highest perceived usability and lowest cognitive load, it also results in the poorest comprehension. In contrast, AI assistance in the \IntermediateAI condition effectively mitigates the potential negative impact of AI on comprehension, with students benefiting from interacting with AI and perceiving this interaction as helpful for learning.  We discuss our results with note-taking, cognitive load, and learning theories to explain the mechanisms behind this trend. We also discuss the implications of designing intermediate AI to support human cognition. 

\subsection{Unpacking the Paradox: Reduced Effort v.s. Enhanced Comprehensions}

This discussion seeks to understand why \IntermediateAI can promote comprehension and support cognitive processes to a certain extent, what mechanisms make it useful, and why \AutomatedAI, despite seeming to ease the burden on students, does not truly aid comprehension and cognitive processes and, in some cases, even hinders them.

Our findings indicate that students in the \IntermediateAI condition scored the highest on both post-lecture and review-modified comprehension tests 
(Table \ref{score}).
The lower performance of the \AutomatedAI condition may be explained by a lack of encoding processes during note-taking \cite{kiewraReviewNotetakingEncodingstorage1989}. 
On the other hand, \IntermediateAI preserved the need for learners to interpret, organize, and adapt information in real-time. This aligns with the encoding-storage paradigm \cite{kiewraReviewNotetakingEncodingstorage1989}, which posits that learning is strengthened when individuals actively process content during note-taking. Besides, \IntermediateAI presents appropriately sized information chunks\cite{tankelevitch2024metacognitive, zhang2025ladica}, enabling students to efficiently scan AI-generated content, evaluate its relevance, and consider how to integrate it into their structure. 

A counterintuitive finding was that even after students used the notes to adjust their responses in the post-test, \AutomatedAI still performed worse. This contrasts with students’ intuitive expectations, as most interviewees reported feeling that structured notes provided the most complete and organized information during the review phase. We believe that this finding aligns with our design rationale, as shown in \S\ref{design consideration}. When AI takes over much of the encoding process, it may effectively reduce the burden of encoding but also increase the challenge of interpreting and making sense of this information during storage. This also ties in with collaborative sense-making theories suggesting that when collaborators largely take over the process of instantiating external representations, it can make consuming encodons more difficult \cite{russellCostStructureSensemaking1993a}. 

Our data show that while overall cognitive load did not differ significantly, the perceived extraneous cognitive load was lowest in the \AutomatedAI condition (Table \ref{table:cognitive_load_anova_means}). 
However, as cognitive load theory suggests, cognitive load is not always detrimental \cite{piolatCognitiveEffortNote2005b, klepsch2017development}. 
Although students reported that interacting with AI in the \IntermediateAI condition required effort, they perceived this effort as enhancing their understanding. \IntermediateAI appears to direct user effort toward meaningful processing rather than mechanical transcription. Our study is aligned with the cognitive load theory in that, in cognitively intensive tasks such as learning and note-taking, the goal of AI assistance should not be minimizing all user effort but rather supporting users in offloading mechanical effort while maintaining space for active reasoning, reflection, and self-monitoring \cite{chen2025we, scott2025does, tankelevitch2024metacognitive}. This perspective aligns with prior work discussing cognitive engagement with AI assistance, which distinguishes cognitive engagement—where users interpret and apply information—from passive engagement, where users simply follow AI suggestions without critical processing \cite{gajos2022people}.

The implications of these findings suggest that we need to be more cautious when providing AI assistance to users, especially on tasks that require substantial cognitive effort. As users naturally prefer to minimize effort, the risk of over-reliance on AI becomes even more pronounced. 
This insight also calls for reconsidering how we measure the effectiveness of AI tools for cognitive tasks. Many studies use task load surveys to measure the amount of load and assume that a lower load is better \cite{hart2006nasa}. However, if these tools are designed for cognitive tasks, such as learning, brainstorming, or meetings, a lower load does not necessarily indicate better outcomes. More appropriate measurements should be considered.

\subsection{Designing Intermediate AI for Real-Time Learning and Note-Taking}
Our study shows that \IntermediateAI significantly promotes student understanding and maintains cognitive engagement during note-taking. The success of \IntermediateAI can be attributed to its design as a “middle ground,” where AI-generated “building blocks” allow students to integrate relevant information into their notes in real-time. Unlike \AutomatedAI, which presents finalized content, \textit{IntermediateAI} invites learners to evaluate, modify, and reorganize the information.
 Rather than reducing effort, \IntermediateAI redistributes it toward meaningful engagement, aligning with theories of desirable difficulty \cite{bjork2011making} and cognitive scaffolding \cite{azevedo2004does}, which suggest that moderate challenge, when well-supported, can enhance comprehension and retention.

However, \IntermediateAI should not be viewed as a one-size-fits-all solution. As our findings show (\S\ref{finding - intent}), students imagined using different types of AI assistance depending on lecture pace, topic complexity, and their own learning goals. 
These various assistance needs align with the Knowledge-Learning-Instruction (KLI) framework \cite{koedinger2012knowledge}, which emphasizes that instructional strategies should match the type of knowledge being acquired. 
While our current implementation of \IntermediateAI supports factual knowledge particularly well—each summary block typically maps to a discrete fact or definition—we observe that students often go beyond simply collecting facts. In our study, learners actively reorganized, grouped, and rephrased building blocks to construct broader conceptual insights or procedural understanding. This reveals both a strength and a design opportunity: \IntermediateAI, by surfacing atomic knowledge units, leaves room for users to engage in deeper knowledge construction through structuring and synthesis. However, to better support learning tasks that involve procedural or conceptual knowledge from the outset, future systems could extend \IntermediateAI to present more connected chunks of content or scaffold cross-turn integration to support comprehension and application better.

In light of these shifting demands, we argue that intermediate AI should be adaptive, not merely in the traditional sense of tailoring scaffolding based on performance assessment \cite{azevedo2004does}, but in a more goal-oriented and engagement-sensitive way. 
Students expressed interest in systems that could infer preferences from their interaction patterns, such as how they restructure summaries, what types of content they drop or emphasize, and how long they hover over certain blocks. While prior models primarily adapt support based on observable learning performance in well-defined problem spaces \cite{kalyuga2009adapting}, our findings highlight the need for AI assistance to adapt in more open-ended learning environments by dynamically adjusting content granularity based on topic complexity, inferring learner state from behavioral signals (e.g., hesitations, deletions), or offering customizable output formats (e.g., bullet points, outlines, concept maps). 

\subsection{Intermediate AI as a Human-AI Collaboration Paradigm}

While our study focuses on the learning context of note-taking, its design implications extend to broader cognitive tasks where AI assists human sensemaking and decision-making. We propose that \textit{Intermediate AI} embodies a distinct paradigm—intermediate AI as a deliberate middleware layer—that supports cognitively intensive work without over-automation.

A key design principle behind \textit{Intermediate AI} is the intentional preservation of the most creative and intellectually demanding aspects of a task for the human user. In our case, \textit{Intermediate AI} alleviates the burden of low-level actions, such as continuous typing, while preserving higher-order sensemaking, allowing learners to evaluate, select, and integrate relevant information \cite{kiewraReviewNotetakingEncodingstorage1989}. Similar human-led constructs appear in other cognitively valuable contexts, such as refining one’s argument during writing \cite{dhillon2024shaping} or sustaining intentionality in meetings \cite{scott2024mental, chen2025we}. In contrast, repetitive or mechanical subtasks—recording, formatting, or retrieval—can be offloaded to AI.

Importantly, \textit{Intermediate AI} is not limited to the format of summarization shown in this paper. It represents a broader design space with multiple representations:  building blocks \cite{hou2025personalized}, embedded questions \cite{chen2025we}, or interactive templates \cite{xu2024jamplate}. 
Our exploration of intermediate AI design in real-time note-taking scenarios further contributes design insights that may generalize to other intermediate AI systems. These include (1) output granularity—how abstract the AI-generated content should be and when to surface it; (2) control dynamics—balancing pull-based user queries with push-based AI suggestions; (3) synchronicity - presenting information chunks in real-time so that people can organize them when they're in their working memory. 

This paradigm is particularly promising in real-time, collaborative settings. 
For example, in group meetings, managing the cognitive demands of listening, reasoning, and contributing simultaneously can be challenging \cite{wangMeetingBridgesDesigning2024}. \textit{Intermediate AI} 
enables proactive AI-generated summaries and user-initiated, in-context search, allowing participants to fluidly engage with support without interrupting their cognitive flow. 
MeetMap similarly adopts this middleware layer by providing a transient holding area for live AI summaries, preventing overload and allowing for timely, contextual interpretation \cite{chen2025meetmap}.

Besides providing real-time AI help, intermediate AI design also opens up opportunities for implicit intent modeling. Users imagine that \textit{Intermediate AI} could progressively learn from their behaviors and use these signals to infer intent and generate more tailored, personalized outputs over time. This idea is echoed in recent system \textsc{Granola} \footnote{Granola: https://www.granola.ai/}, which observes light-touch user inputs during meetings and generates personalized post-meeting summaries aligned with the input.

In sum, much of today’s AI development centers on task automation—optimizing speed, output, and scale. While effective, this also reduces opportunities for learning and critical thinking.
We advocate for \textit{Intermediate AI} as a design paradigm that re-centers human agency, reducing unnecessary effort while preserving the right kind of effort—effort that enables engagement, reflection, and ownership. 
We hope to elicit further discussion on designing AI not only for productivity but also for thoughtfulness and augmenting human cognition.

\subsection{Limitation}
\label{limitation}
We acknowledge the limitations of this work. 1) Our sample size was limited, preventing us from using more complex models to explore relationships between variables.  
2) The videos used in this study were only ten minutes long, which may not fully capture the impact of AI assistance over extended lectures or in real, long-term courses. Future work could address this with a larger sample, deploying a real classroom setting.
Although we controlled for video difficulty and ensured the content was accessible without prior knowledge, our participants were from different academic years with varying knowledge levels. This may have affected their final scores; however, we mitigated this by treating individual differences as a random effect in our model, which likely minimized its influence. Future research could incorporate pre- and post-tests to measure learning gains more accurately. Our current analysis of user interactions with the dropped AI notes only compares the final state of the notes to the original AI-generated content. Although we logged each user interaction with the editor (e.g., typing events during the session), fully capturing the dynamic editing process — such as incremental additions, deletions, or reorganizations of AI content — would require more detailed data logging and semantic analysis of intermediate note states. We will consider logging this process data and further analyzing how users interact with the AI-generated content in real-time in future studies.

\section{Conclusion}
As AI tools become integral to tasks requiring high cognitive effort, such as note-taking, questions arise about their effects on cognitive engagement. This paper explores the "AI Assistance Dilemma" in note-taking by studying the impact of different AI support levels on user engagement and comprehension through a within-subject experiment where participants (N=30) took notes during lecture videos under three conditions: \AutomatedAI (high assistance with structured notes), \IntermediateAI (moderate assistance with real-time summary), and \MinimalAI (low assistance with transcript); results indicate that Intermediate AI led to the highest post-test scores, while Automated AI resulted in the lowest, although participants favored Automated AI for its perceived ease and lower cognitive effort. The results highlight a potential mismatch between preferred convenience and cognitive benefits. Our study provides insights into designing AI systems that maintain cognitive engagement, with implications for developing moderate AI support in cognitive tasks.

\section*{Acknowledgments}

This material is based upon work supported by the National Science Foundation under Grant Number IIS-2302564.

\bibliographystyle{ACM-Reference-Format}
\bibliography{reference}


\end{document}